\documentclass[12pt]{iopart}
\input{epsf}

\begin{document}

\article{Strangeness in Quark Matter 2003}{Strangeness from 
$20 \, A \rm GeV$ to $158 \, A \rm GeV$}

\author{Volker Friese}

\address{Gesellschaft f{\"u}r Schwerionenforschung \\
Planckstr.~1, D-64291 Darmstadt, Germany}

\ead{v.friese@gsi.de}

\author{for the NA49 collaboration:}

\author{ \small
C.~Alt$^{9}$, T.~Anticic$^{19}$, B.~Baatar$^{8}$, D.~Barna$^{4}$,
J.~Bartke$^{6}$,  M.~Behler$^{13}$,
L.~Betev$^{9}$, H.~Bia{\l}\-kowska$^{17}$, A.~Billmeier$^{9}$,
C.~Blume$^{7}$,  B.~Boimska$^{17}$, M.~Botje$^{1}$,
J.~Bracinik$^{3}$, R.~Bramm$^{9}$, R.~Brun$^{10}$,
P.~Bun\v{c}i\'{c}$^{9,10}$, V.~Cerny$^{3}$, 
P.~Christakoglou$^{2}$, O.~Chvala$^{15}$,
J.G.~Cramer$^{16}$, P.~Csat\'{o}$^{4}$, P.~Dinkelaker$^{9}$,
V.~Eckardt$^{14}$, P.~Filip$^{14}$,
H.G.~Fischer$^{10}$, Z.~Fodor$^{4}$, P.~Foka$^{7}$, P.~Freund$^{14}$,
V.~Friese$^{7,13}$, J.~G\'{a}l$^{4}$,
M.~Ga{\'z}dzicki$^{9}$, G.~Georgopoulos$^{2}$, E.~G{\l}adysz$^{6}$, 
S.~Hegyi$^{4}$, C.~H\"{o}hne$^{13}$, 
K.~Kadija$^{19}$, A.~Karev$^{14}$, S.~Kniege$^{9}$,
V.I.~Kolesnikov$^{8}$, T.~Kollegger$^{9}$, 
R.~Korus$^{12}$, M.~Kowalski$^{6}$, 
I.~Kraus$^{7}$, M.~Kreps$^{3}$, M.~van~Leeuwen$^{1}$, 
P.~L\'{e}vai$^{4}$, A.I.~Malakhov$^{8}$, 
C.~Markert$^{7}$, B.W.~Mayes$^{11}$, G.L.~Melkumov$^{8}$,
C.~Meurer$^{9}$,
A.~Mischke$^{7}$, M.~Mitrovski$^{9}$, 
J.~Moln\'{a}r$^{4}$, St.~Mr{\'o}wczy{\'n}ski$^{12}$,
G.~P\'{a}lla$^{4}$, A.D.~Panagiotou$^{2}$,
K.~Perl$^{18}$, A.~Petridis$^{2}$, M.~Pikna$^{3}$, L.~Pinsky$^{11}$,
F.~P\"{u}hlhofer$^{13}$,
J.G.~Reid$^{16}$, R.~Renfordt$^{9}$, W.~Retyk$^{18}$,
C.~Roland$^{5}$, G.~Roland$^{5}$, 
M. Rybczy{\'n}ski$^{12}$, A.~Rybicki$^{6,10}$,
A.~Sandoval$^{7}$, H.~Sann$^{7}$, N.~Schmitz$^{14}$, P.~Seyboth$^{14}$,
F.~Sikl\'{e}r$^{4}$, B.~Sitar$^{3}$, E.~Skrzypczak$^{18}$,
G.~Stefanek$^{12}$,
 R.~Stock$^{9}$, H.~Str\"{o}bele$^{9}$, T.~Susa$^{19}$,
I.~Szentp\'{e}tery$^{4}$, J.~Sziklai$^{4}$,
T.A.~Trainor$^{16}$, D.~Varga$^{4}$, M.~Vassiliou$^{2}$,
G.I.~Veres$^{4}$, G.~Vesztergombi$^{4}$,
D.~Vrani\'{c}$^{7}$, S.~Wenig$^{10}$, A.~Wetzler$^{9}$,
Z.~W{\l}odarczyk$^{12}$
I.K.~Yoo$^{13}$, J.~Zaranek$^{9}$, J.~Zim\'{a}nyi$^{4}$
}

\vspace{0.5cm}

\address{$^{1}$ NIKHEF, Amsterdam, Netherlands}
\address{$^{2}$Department of Physics, University of Athens, Athens, Greece}
\address{$^{3}$Comenius University, Bratislava, Slovakia}
\address{$^{4}$KFKI Research Institute for Particle and Nuclear Physics, 
Budapest, Hungary}
\address{$^{5}$MIT, Cambridge, USA}
\address{$^{6}$Institute of Nuclear Physics, Cracow, Poland}
\address{$^{7}$Gesellschaft f\"{u}r Schwerionenforschung (GSI), 
Darmstadt, Germany}
\address{$^{8}$Joint Institute for Nuclear Research, Dubna, Russia}
\address{$^{9}$Fachbereich Physik der Universit\"{a}t Frankfurt, 
Frankfurt, Germany}
\address{$^{10}$CERN, Geneva, Switzerland}
\address{$^{11}$University of Houston, Houston, TX, USA}
\address{$^{12}$Institute of Physics {\'S}wi{\,e}tokrzyska Academy, 
Kielce, Poland}
\address{$^{13}$Fachbereich Physik der Universit\"{a}t Marburg, 
Marburg, Germany}
\address{$^{14}$Max-Planck-Institut f\"{u}r Physik, Munich, Germany}
\address{$^{15}$Institute of Particle and Nuclear Physics, Charles 
University, Prague, Czech Republic}
\address{$^{16}$Nuclear Physics Laboratory, University of Washington, 
Seattle, WA, USA}
\address{$^{17}$Institute for Nuclear Studies, Warsaw, Poland}
\address{$^{18}$Institute for Experimental Physics, University of 
Warsaw, Warsaw, Poland}
\address{$^{19}$Rudjer Boskovic Institute, Zagreb, Croatia}

\begin{abstract}
New results from the energy scan programme of NA49, in particular kaon 
production at $30 \, A \rm GeV$ and $\phi$ production at $40$ and 
$80 \, A \rm GeV$ are presented. The $K^+/\pi^+$ ratio shows a pronounced 
maximum at $30 \, A \rm GeV$; the kaon slope parameters are constant at 
SPS energies. Both findings support the scenario of a phase transition 
at about $30 \, A \rm GeV$ beam energy. The $\phi/\pi$ ratio increases smoothly 
with beam energy, showing an energy dependence similar to $K^-/\pi^-$. The 
measured particle yields can be reproduced by a hadron gas model, with 
chemical freeze-out parameters on a smooth curve in the $T-\mu_B$ plane. 
The transverse spectra can be understood as resulting from a rapidly 
expanding, locally equilibrated source. No evidence for an earlier kinetic 
decoupling of heavy hyperons is found.
\end{abstract}




\section{Introduction}
Strange particle production provides an important tool to understand the 
reaction dynamics of relativistic heavy-ion collisions.  The measured 
yields at top SPS energy can be interpreted as the 
result of the decay of a coherent prehadronic state, filling the available 
hadronic phase space. Several indications suggest that this state is indeed 
a deconfined phase of matter \cite{stock}. In order to search for onset 
phenomena signalling a phase transition at lower collision energies,
the NA49 collaboration proposed
an energy scan programme for central Pb+Pb collisions, spanning the beam 
energies between top AGS ($11.7 \, A \rm GeV$) and top SPS 
($158 \, A \rm GeV$). This programme was completed in 2002 with data 
taking at $20$ and $30 \, A \rm GeV$. In this contribution, we will focus 
on preliminary results on kaons at $30 \, A \rm GeV$ and $\phi$ mesons at 
$40$ and $80 \, A \rm GeV$. NA49 results on the energy dependence of 
$\Lambda$ production were shown recently \cite{na49_lambda}; preliminary 
data of $\Xi$ and $\Omega$ were presented in this conference 
\cite{sqm03_meurer,sqm03_mitrov}.

\begin{table}[bh]
\caption{\label{tab:datasets} Data sets taken in the NA49 energy scan 
programme. For the analysis at $158 \, A \rm GeV$, the $5 \, \%$ most
central events were selected offline.}
\begin{indented}
\item[]\begin{tabular}{@{}cccccc}
\br
$E_{beam} \, [A\rm{GeV}]$  &  $\sqrt{s} \, [A\rm{GeV}]$  & $y_{cm}$ 
& Centrality  &  Event Statistics  &  Run period \\
\mr
158  &  17.3  & 2.91 & $10 \, \%$  &   800 k   &  1996  \\
158  &  17.3  & 2.91 & $20 \, \%$  & 3,000 k   &  2000  \\
 80  &  12.3  & 2.57 & $7 \, \%$   &   300 k   &  2000  \\
 40  &   8.8  & 2.22 & $7 \, \%$   &   700 k   &  1999  \\
 30  &   7.6  & 2.08 & $7 \, \%$   &   400 k   &  2002  \\
 20  &   6.4  & 1.88 & $7 \, \%$   &   300 k   &  2002  \\
\br
\end{tabular}
\end{indented}
\end{table}

\section{Experiment and data sets}
The NA49 detector \cite{nim}, operating in fixed-target mode at the 
CERN-SPS, consists mainly of four large tracking chambers 
backed up with TOF scintillator walls. 
Momentum determination is provided by tracking in the field of two 
superconducting magnets; the collision centrality is determined from
the measurement of the 
energy deposited in a zero-degree calorimeter. Particle identification is 
achieved by the measurements of the specific energy loss $dE/dx$ 
(resolution $\approx 4 \, \%$) in the TPCs and the time-of-flight 
(resolution $\approx 60 \, \rm ps$) with the scintillator walls. The magnetic 
field was scaled proportional to the beam energy to achieve similar
geometrical acceptances. Table \ref{tab:datasets} summarises 
the data sets taken within the energy scan programme. The data at 
$20 \, A \rm GeV$ are not yet analysed.


\section{Kaon and pion production}

\begin{figure}[b]
\begin{center}
\begin{minipage} {5cm}
\begin{center}
\epsfxsize=5cm
\epsfbox{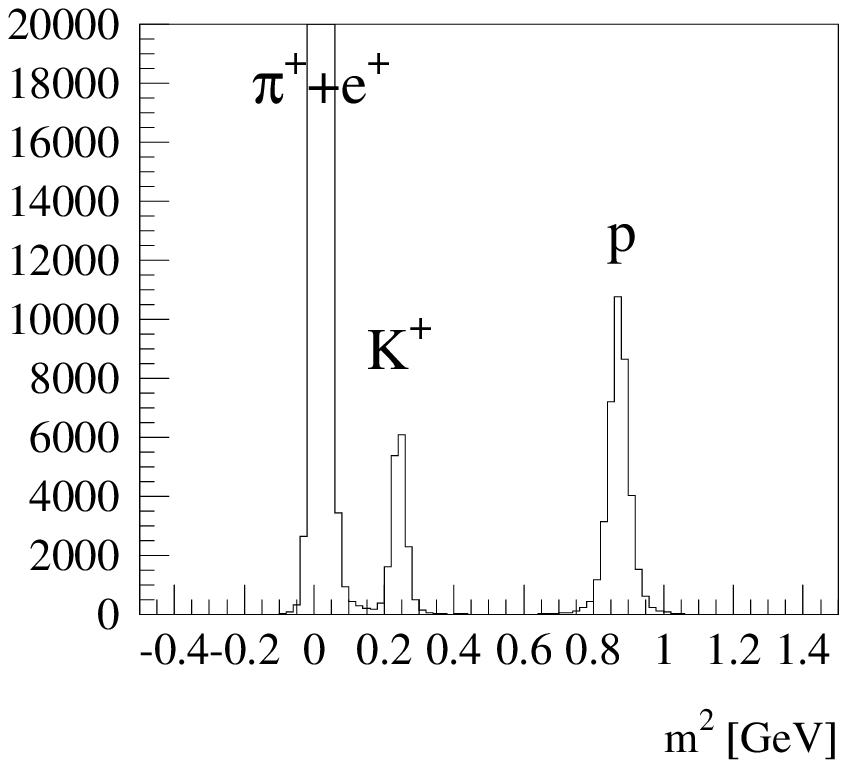}
\end{center}
\end{minipage}
\begin{minipage} {5cm}
\begin{center}
\epsfxsize=5cm
\epsfbox{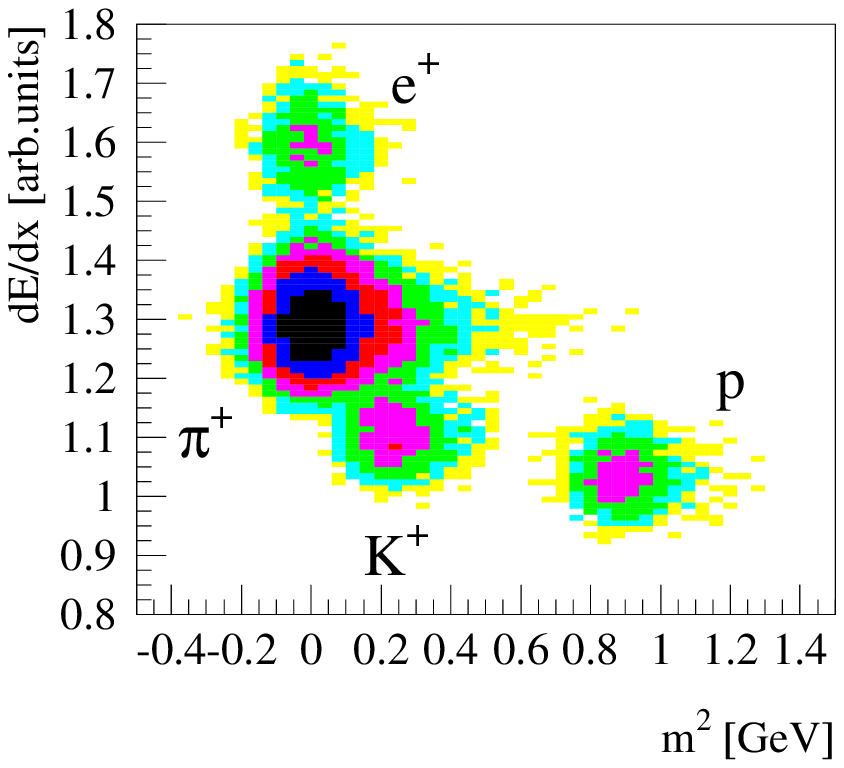}
\end{center}
\end{minipage}
\begin{minipage} {5cm}
\begin{center}
\epsfxsize=5cm
\epsfbox{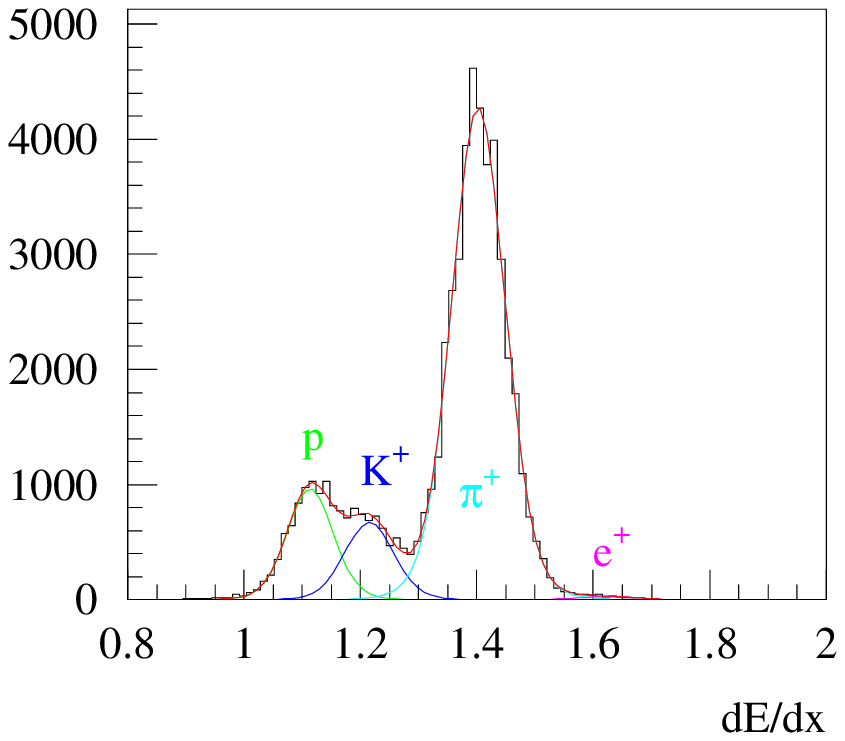}
\end{center}
\end{minipage}
\end{center}
\vspace{-0.5cm}
\caption{\label{fig:kaonid} Examples for kaon identification in NA49
at $40 \, A \rm GeV$. 
Left: TOF ($p=2 \, \rm GeV$); centre: TOF+$dE/dx$ ($p=6 \, \rm GeV$); 
right: $dE/dx$ ($p = 15 \, \rm GeV$).}
\end{figure}

In addition to the published results for $40$, $80$ and 
$158 \, A \rm GeV$ \cite{energy_paper}, we have recently analysed 
kaon and pion production at $30 \, A \rm GeV$. Depending on their 
momentum, charged kaons were identified by TOF, combined TOF and 
$dE/dx$ or $dE/dx$ alone, as figure \ref{fig:kaonid} illustrates. 
GEANT calculations were used to correct the raw yields for geometrical 
acceptance, in-flight decay and the efficiency of the time-of-flight 
system. The losses due to track reconstruction inefficiencies were 
studied by embedding simulated tracks into real raw data events and 
reconstructing them with the standard software. 
$\pi^-$ yields were obtained from the distributions of 
all negatively charged hadrons by subtraction of the contributions from 
$K^-$, $\bar p$, $e^-$ and secondary hadrons from weak decays and 
interactions in the detector. The $\pi^+$ were not measured directly
but calculated from the $\pi^-$ yield assuming that the $\pi^+ / \pi^-$ ratio,
which was measured in regions where both TOF and $dE/dx$ information
were available, is constant over phase space. Details of the kaon and 
pion analyses are given in \cite{energy_paper}. 

\begin{figure}[t]
\begin{center}
\begin{minipage} {5cm}
\begin{center}
\epsfxsize=5cm
\epsfbox{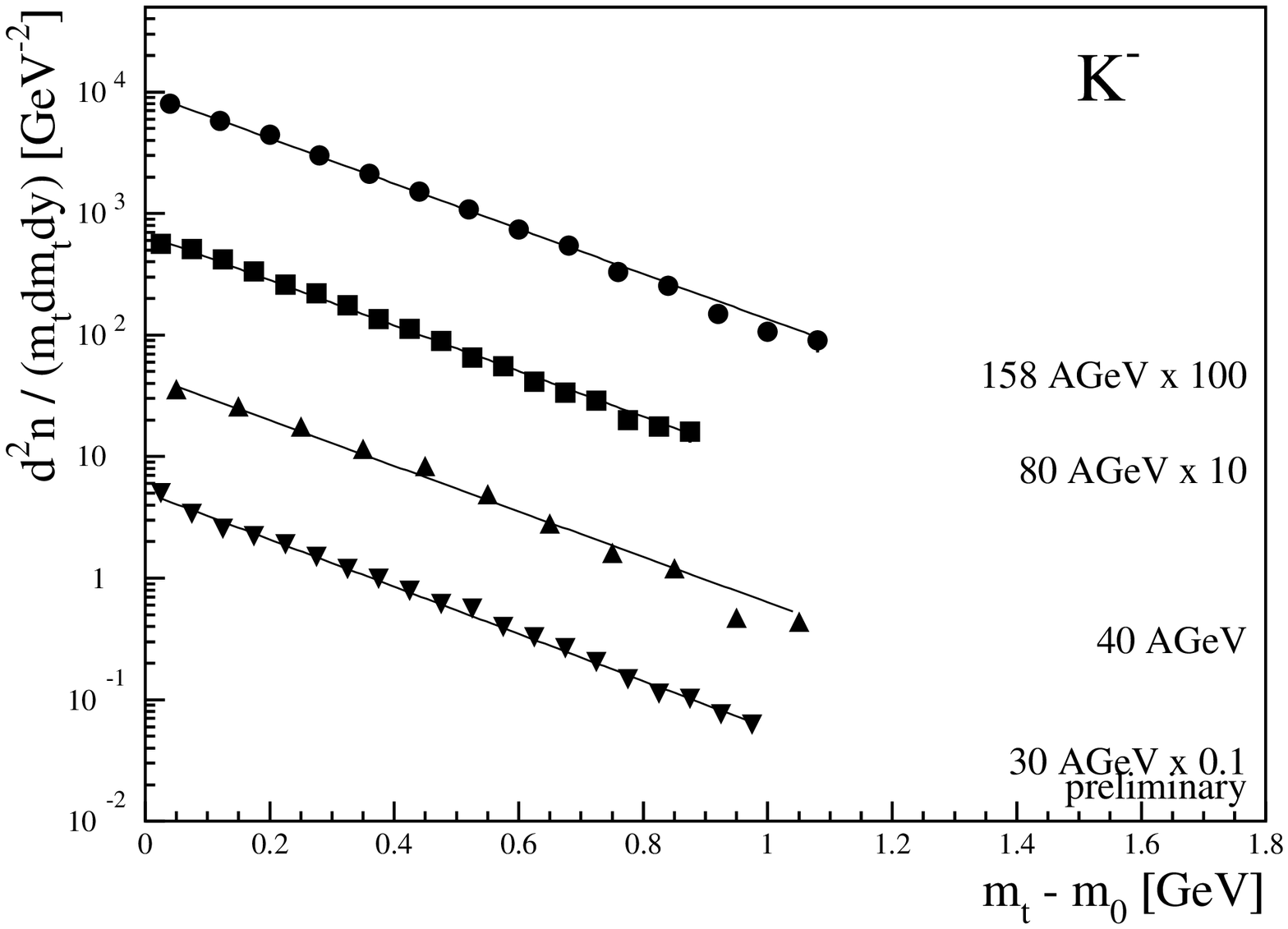}
\end{center}
\end{minipage}
\begin{minipage} {5cm}
\begin{center}
\epsfxsize=5cm
\epsfbox{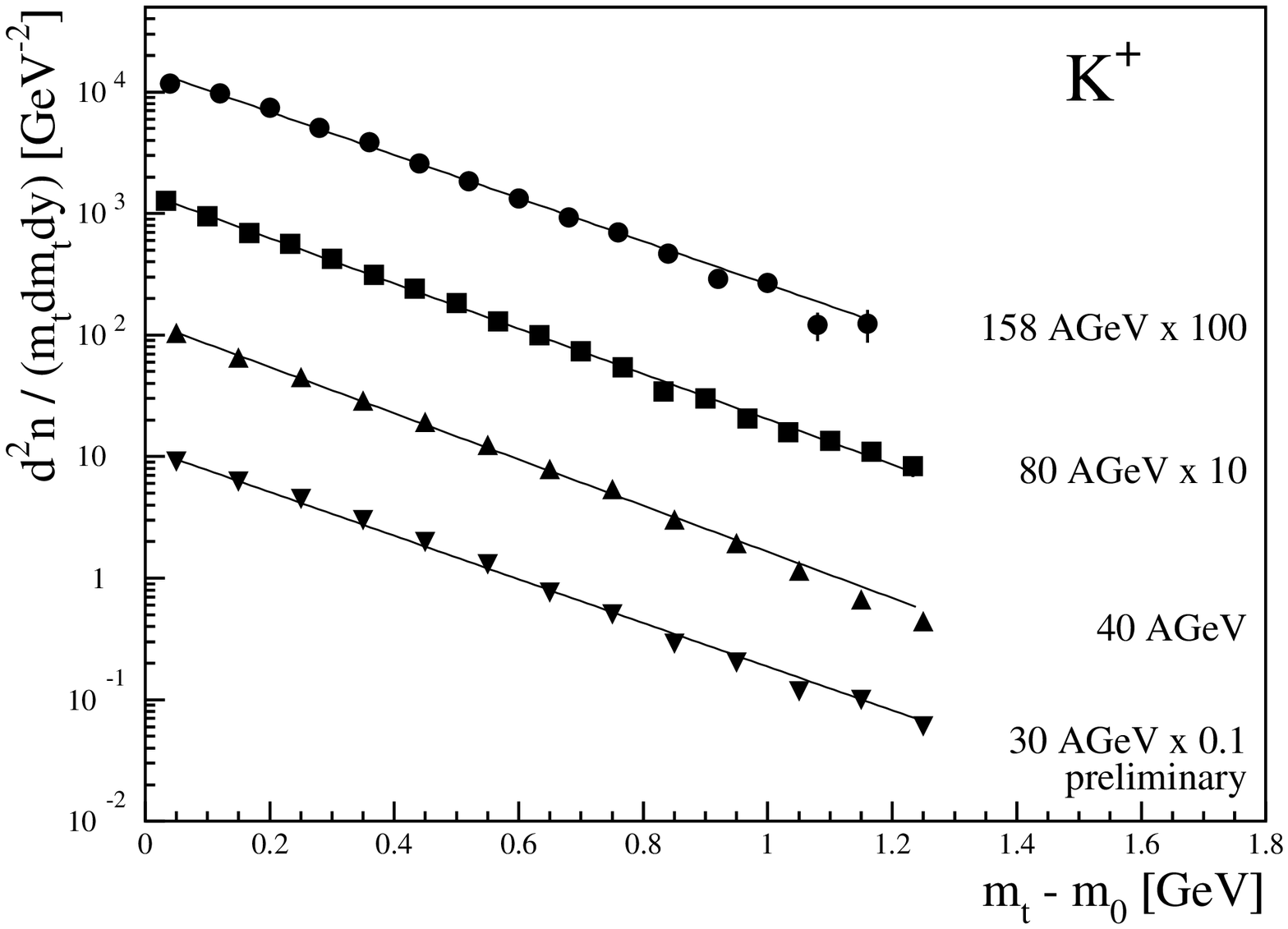}
\end{center}
\end{minipage}
\end{center}
\caption{\label{fig:kaon_mtspec} Transverse spectra at midrapidity for 
$K^-$ and $K^+$, measured in the time-of-flight system. The spectra for 
the different beam energies are scaled for better visibility.}
\end{figure}

\begin{figure}[h]
\begin{center}
\begin{minipage} {5cm}
\begin{center}
\epsfxsize=5cm
\epsfbox{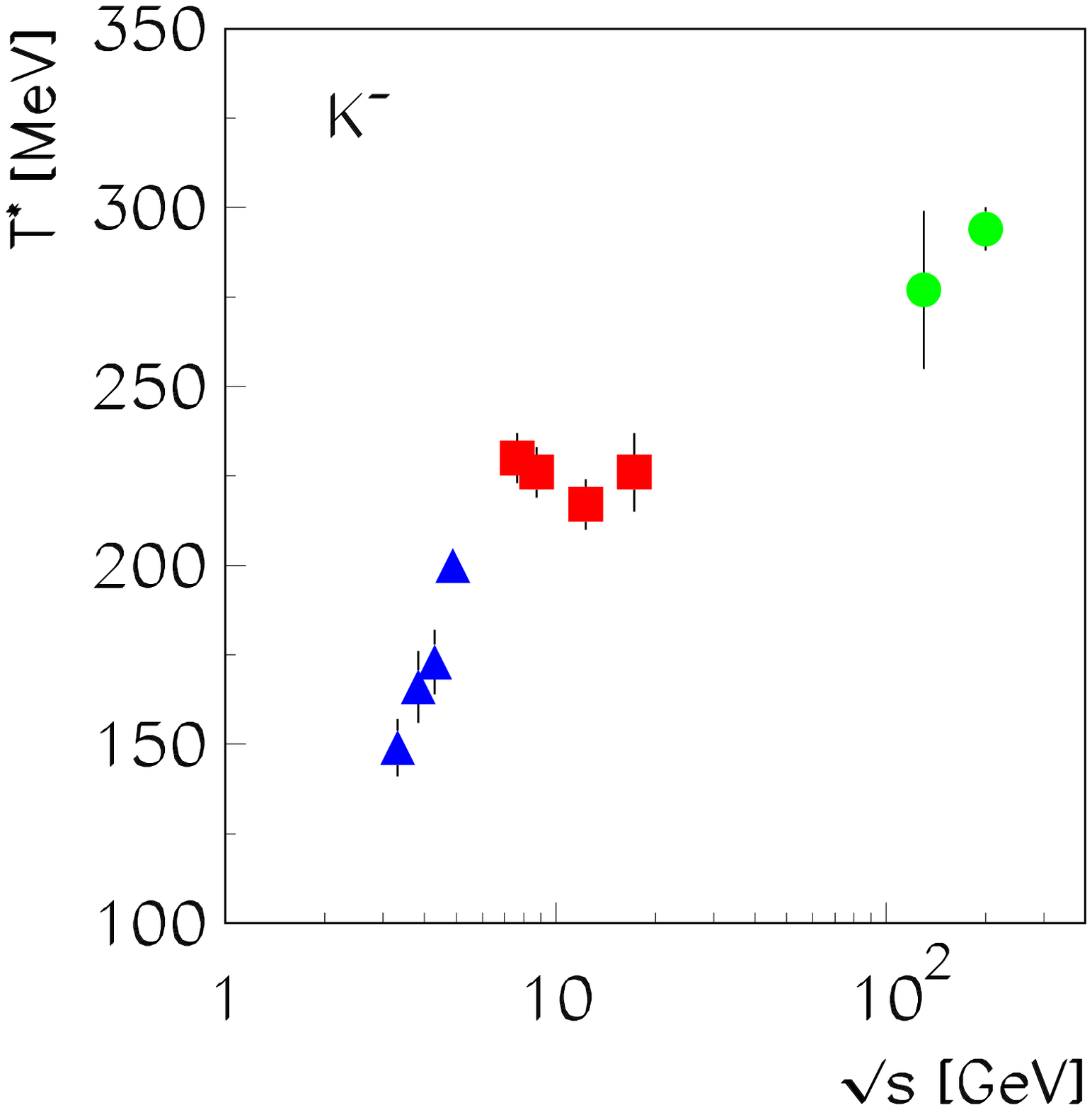}
\end{center}
\end{minipage}
\begin{minipage} {5cm}
\begin{center}
\epsfxsize=5cm
\epsfbox{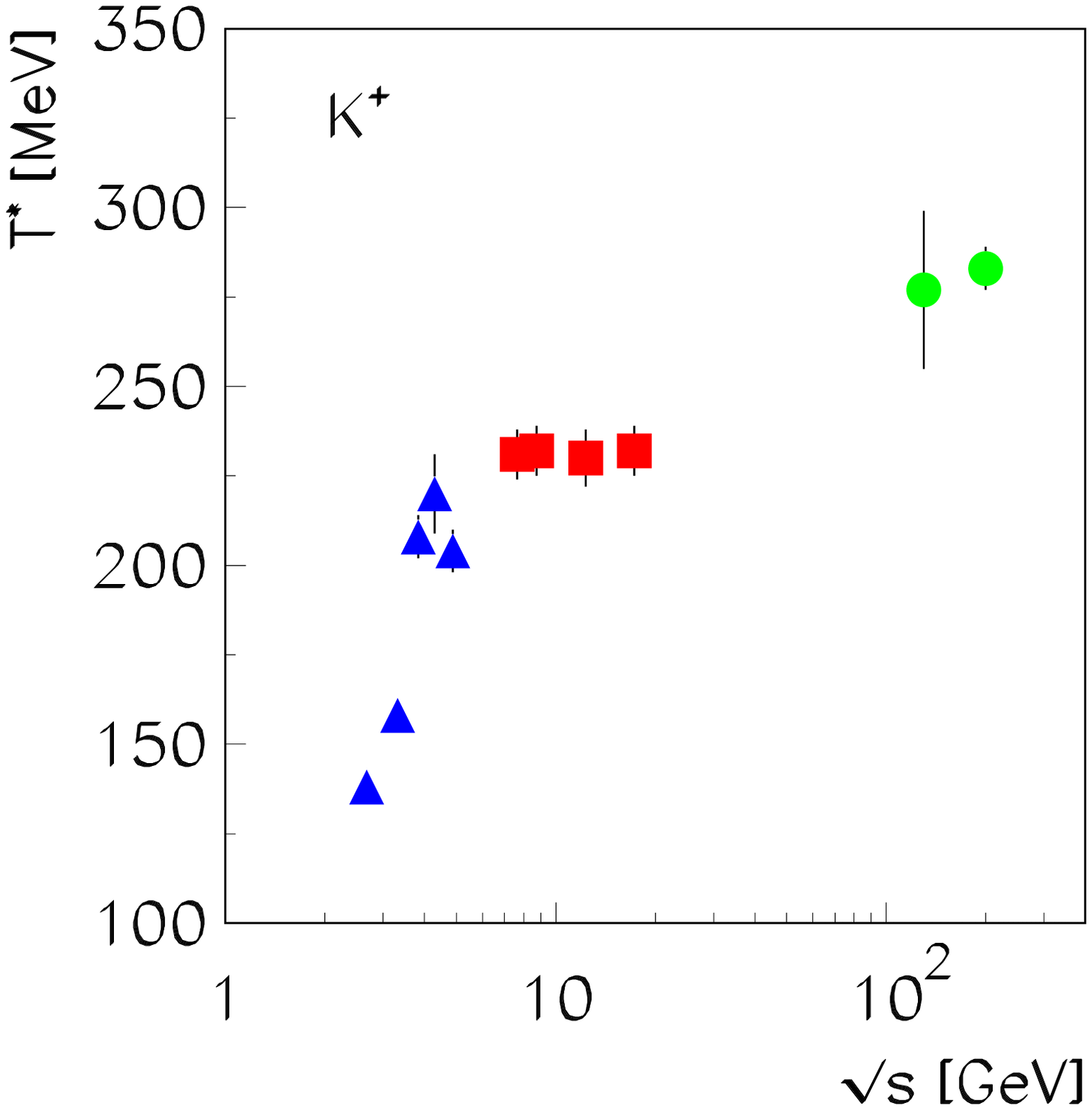}
\end{center}
\end{minipage}
\end{center}
\vspace{-0.5cm}
\caption{\label{fig:kaonslopes} Slope parameters for $K^+$ and $K^-$ as 
measured by NA49, compared to measurements at AGS and RHIC 
(for references see \cite{energy_paper}), as functions of 
collision energy.}
\end{figure}

The newly obtained midrapidity $m_t$ spectra of $K^+$ and $K^-$ at 
$30 \, A \rm GeV$, measured in the TOF system, are compared in figure 
\ref{fig:kaon_mtspec} to those at higher beam energies. At all energies, 
the spectra exhibit an exponential shape to very good accuracy. Moreover, a 
thermal fit of the form $\case {\rmd N} {m_t \rmd m_t \rmd y} 
\propto \rme^{-m_t/T}$ yields almost identical slope parameters 
for all beam energies as demonstrated in figure \ref{fig:kaonslopes}. 
The excitation function of $T$ shows a steep rise for AGS energies,
stays constant over the complete SPS energy range and rises again towards the 
values measured at RHIC. This surprising result points towards a softening
of the equation of state at SPS energies as discussed in \cite{gorgad03}.

\begin{figure}[t]
\begin{center}
\begin{minipage} {5cm}
\begin{center}
\epsfxsize=5cm
\epsfbox{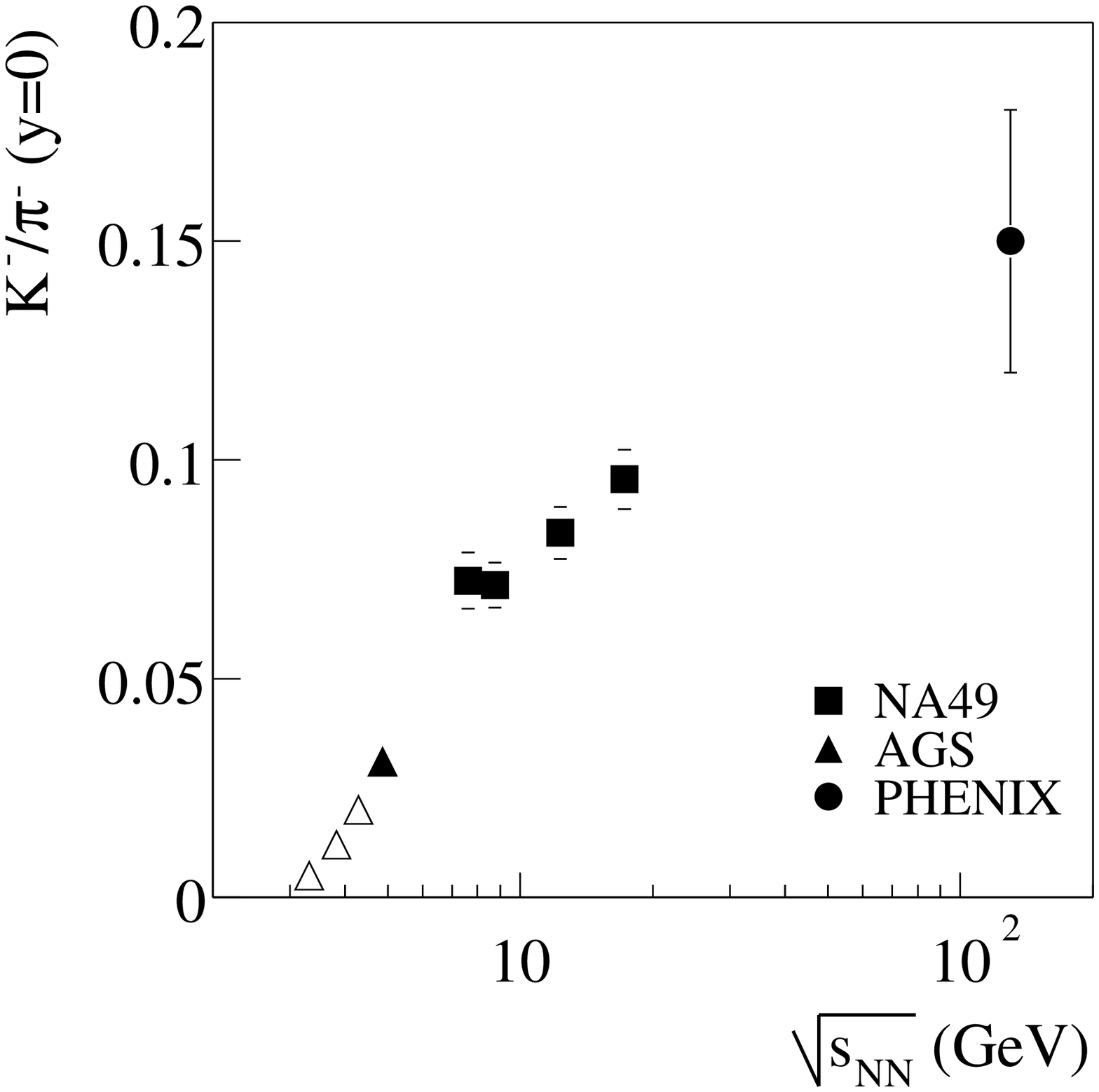}
\end{center}
\end{minipage}
\begin{minipage} {5cm}
\begin{center}
\epsfxsize=5cm
\epsfbox{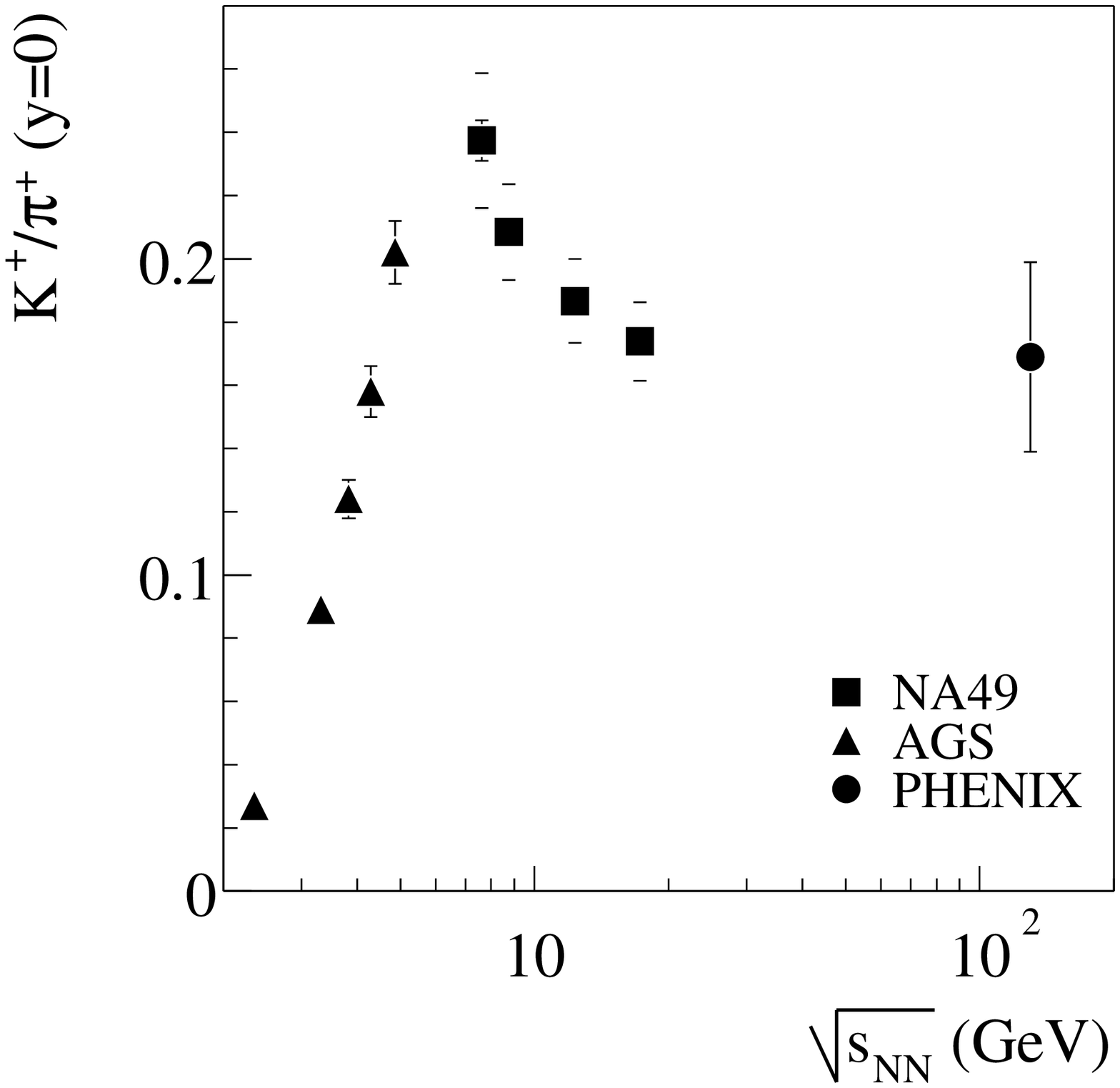}
\end{center}
\end{minipage}
\begin{minipage} {5cm}
\begin{center}
\epsfxsize=5cm
\epsfbox{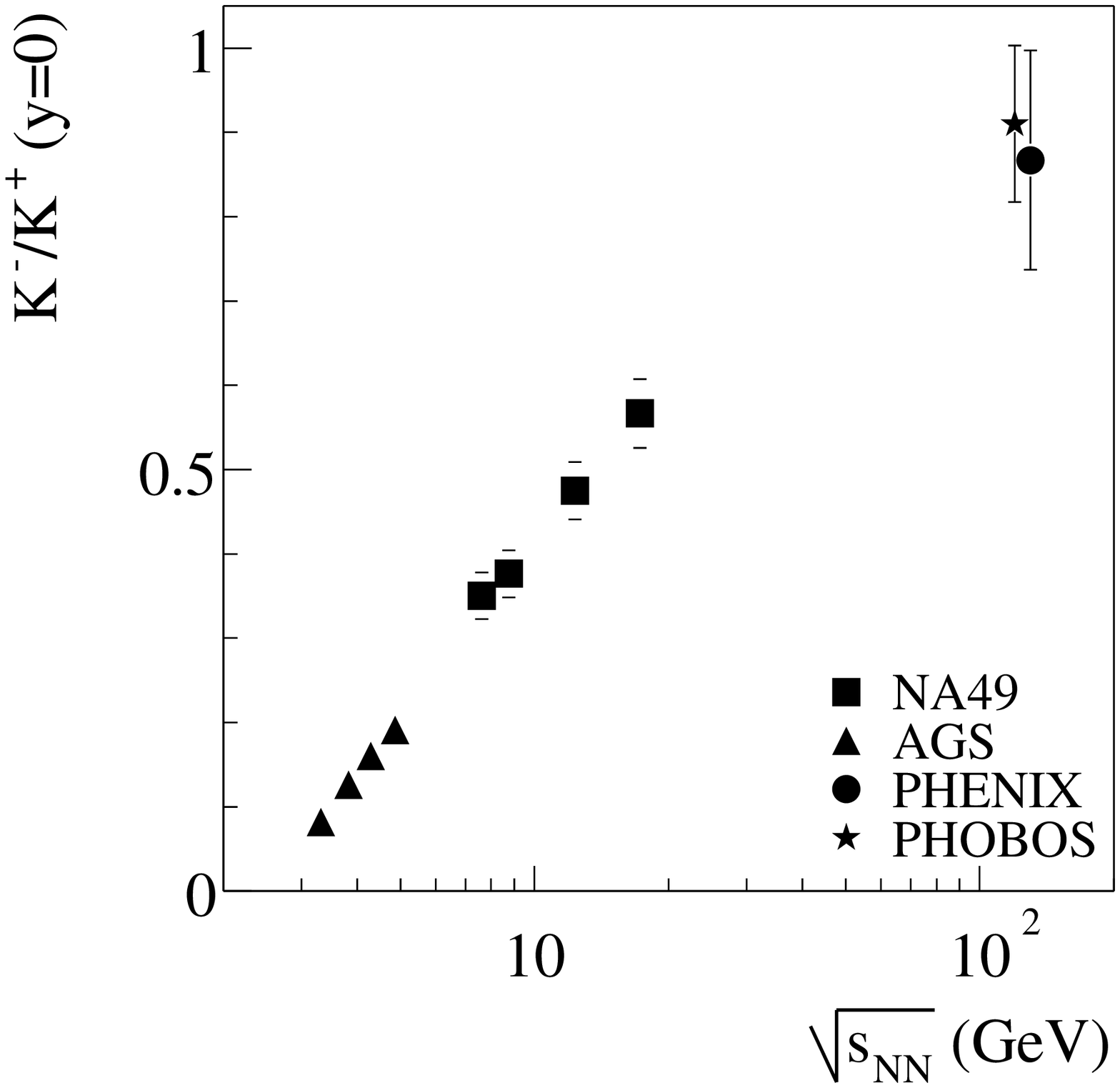}
\end{center}
\end{minipage}
\end{center}
\caption{\label{fig:kpi_midrap} Midrapidity yield ratios as functions 
of collision energy. References for the measurements at AGS and RHIC can 
be found in \cite{energy_paper}.}
\end{figure}

Extrapolating the measured yields to full $p_t$ by the exponential fit 
function, we obtained the midrapidity yield ratios as plotted in figure 
\ref{fig:kpi_midrap}. The $K^+/\pi^+$  ratio at $30 \, A \rm GeV$ 
confirms the previously observed trend of a decrease in this ratio with 
beam energy at SPS \cite{energy_paper}. The $K^-/\pi^-$ ratio seems to also 
show a small irregularity at $30 \, A \rm GeV$. In contrast, no anomalous 
behaviour is seen in the $K^-/K^+$ ratio.

\begin{figure}
\begin{center}
\epsfxsize=10cm
\epsfbox{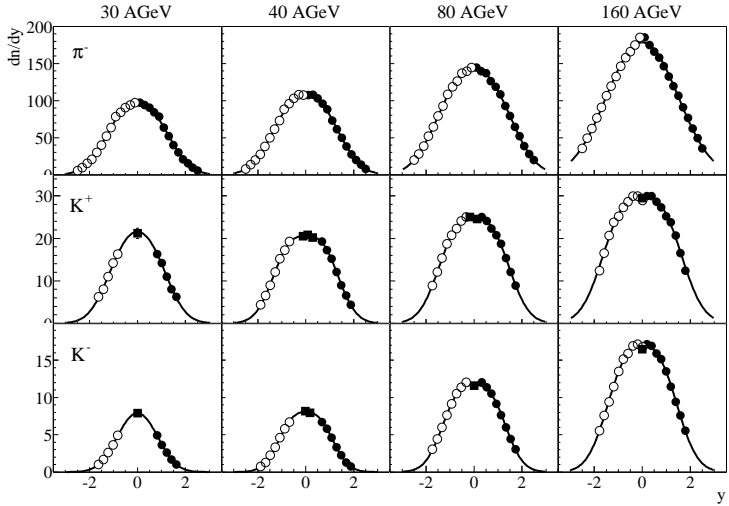}
\vspace{-0.5cm}
\caption{\label{fig:kaon_rapidity} Rapidity distributions of $\pi^-$, $K^+$ 
and $K^-$ for four different beam energies. Filled symbols are measured, 
open ones are reflected at midrapidity. Squares denote the TOF 
measurements, circles the $dE/dx$ results.}
\end{center}
\end{figure}

The rapidity spectra of charged kaons and $\pi^-$ as measured by NA49 are 
shown in figure \ref{fig:kaon_rapidity}. They were parametrised by the 
sum of two Gaussians displaced symmetrically with respect to midrapidity 
as indicated by the full lines. We observe an increase of the 
width of the rapidity distributions with beam energy for all three 
particle types. 

\begin{figure}[t]
\begin{center}
\begin{minipage} {5cm}
\begin{center}
\epsfxsize=5cm
\epsfbox{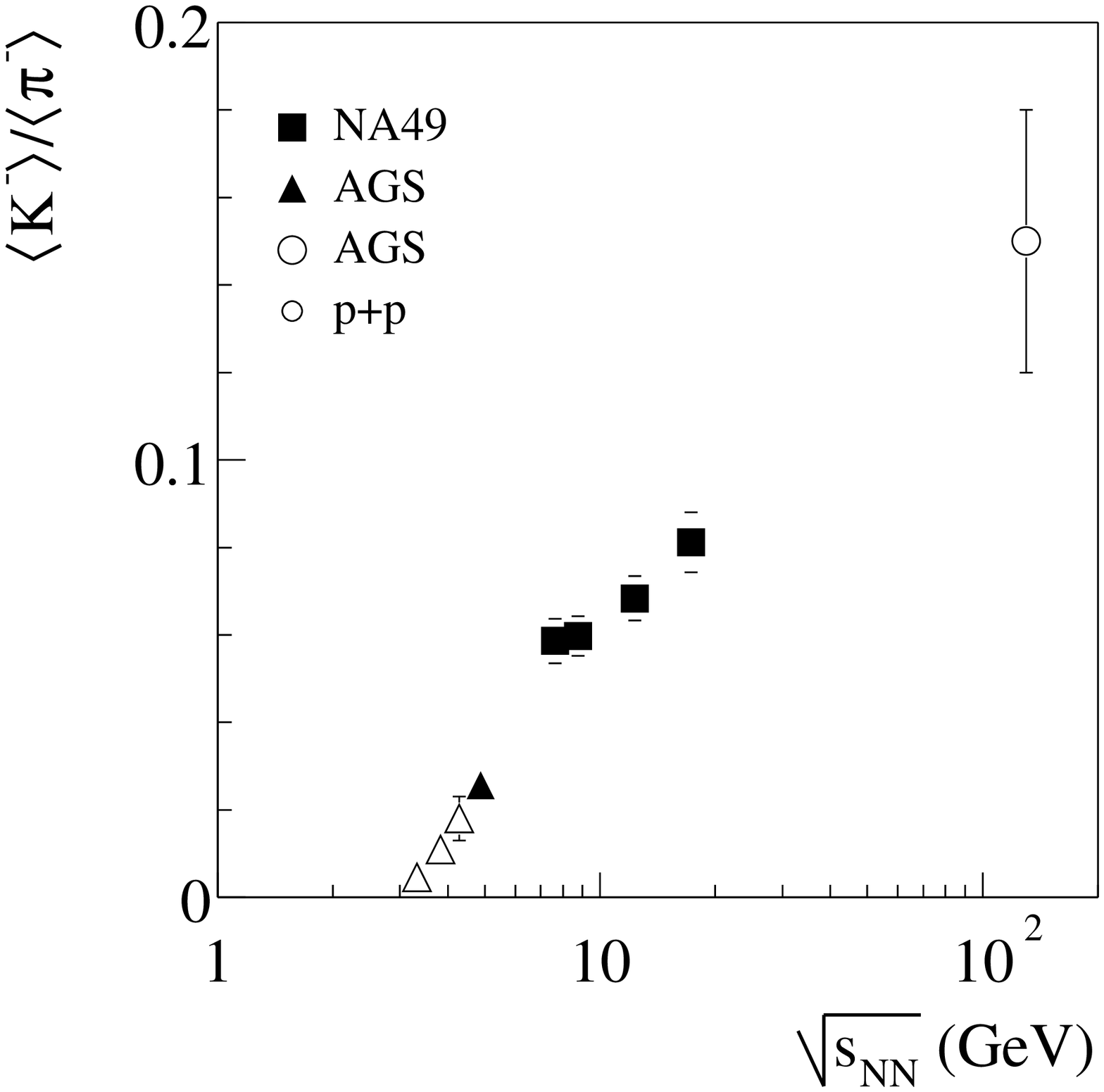}
\end{center}
\end{minipage}
\begin{minipage} {5cm}
\begin{center}
\epsfxsize=5cm
\epsfbox{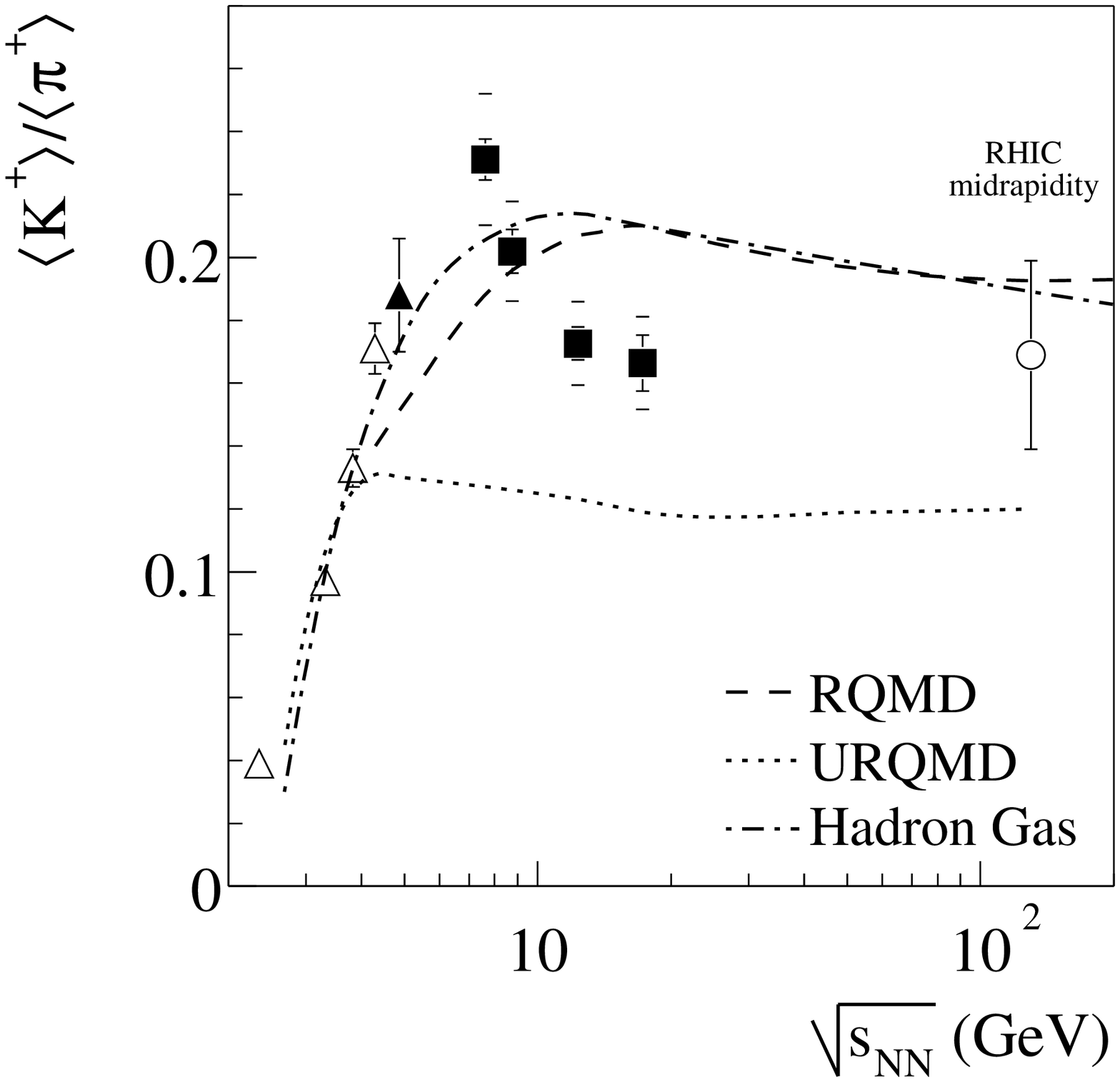}
\end{center}
\end{minipage}
\begin{minipage} {5cm}
\begin{center}
\epsfxsize=5cm
\epsfbox{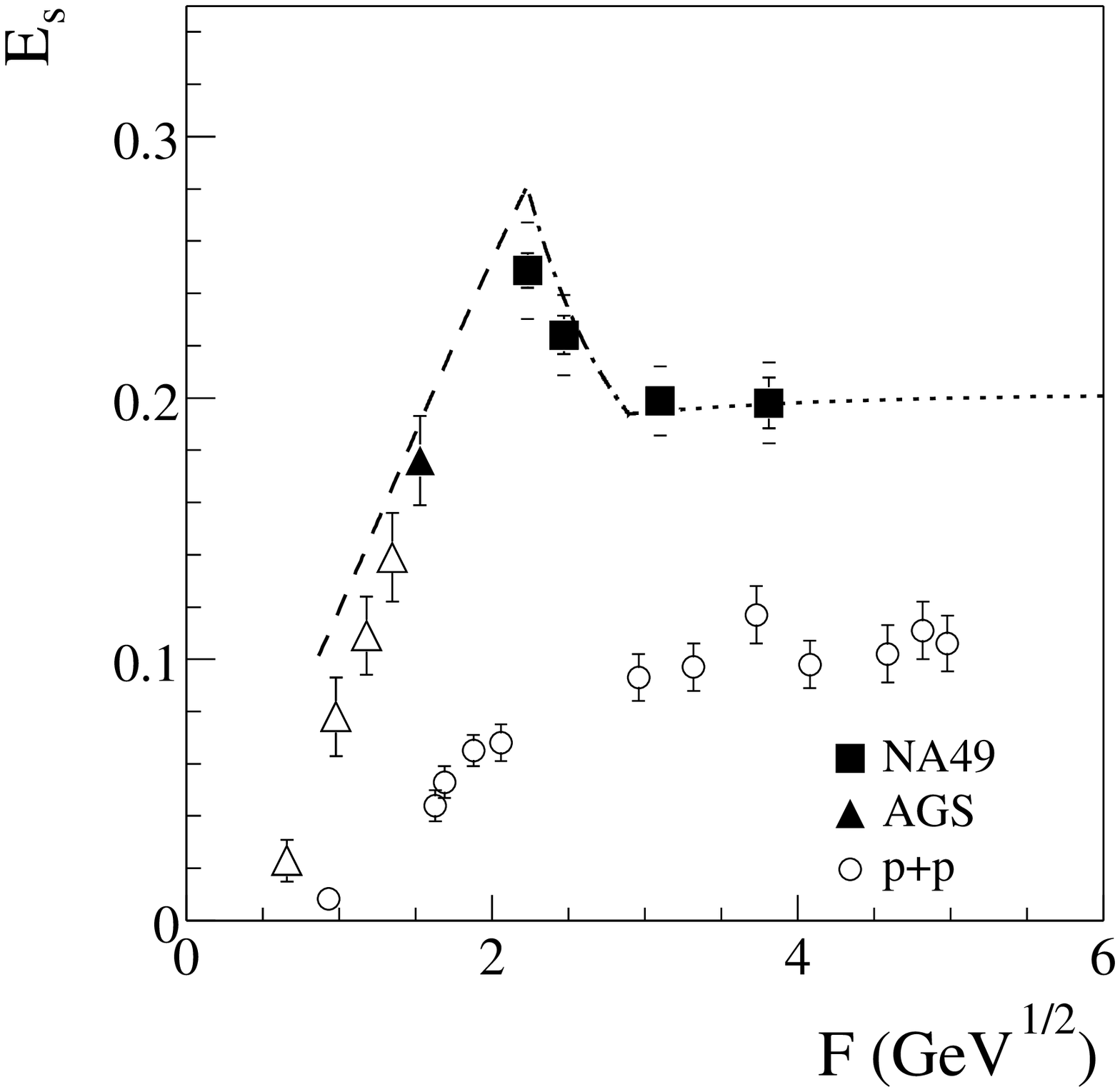}
\end{center}
\end{minipage}
\end{center}
\vspace{-0.5cm}
\caption{\label{fig:kpi_total} $K/\pi$ ratios in full phase space as 
functions of collision energy. Left: $K^-/\pi^-$; centre: $K^+/\pi^+$, 
compared to the predictions of RQMD \cite{rqmd}, UrQMD \cite{urqmd} 
and the extended hadron gas model \cite{pbm}; 
right: Strangeness-to-pion ratio $E_S = \case {\langle \Lambda \rangle
 + \langle K + \bar{K} \rangle} {\langle \pi \rangle}$, 
compared to the prediction of the SMES \cite{SMES}. References 
for data not measured by NA49 can be found in \cite{energy_paper}.}
\end{figure}

Total yields were obtained by integrating the fits to the measured 
rapidity distributions. 
Comparing figure \ref{fig:kpi_total} to figure \ref{fig:kpi_midrap},
it can be seen that the energy dependencies of the $K/\pi$ ratios
in full phase space are similar to those at midrapidity.
A sharp maximum is observed at $30 \, A \rm GeV$ in 
$K^+/\pi^+$, which is not reproduced by either the extended hadron gas model 
\cite{pbm} or transport codes \cite{rqmd,urqmd}. 
It should be noted that this behaviour is not 
seen in p+p collisions. On the other 
hand, a spiky feature in the excitation function of the total strangeness 
to pion ratio $E_S$ was predicted by the Statistical Model of The Early 
Stage \cite{SMES}, assuming a phase transition from confined matter to a 
Quark-Gluon-Plasma at low SPS energies. As figure \ref{fig:kpi_total} 
demonstrates, the preliminary data at $30 \, A \rm GeV$ together with
the measurements at higher SPS energies are in good agreement with the 
predictions of this model. Note that the error bars for the NA49 results 
are dominated by systematic uncertainties which are to a large extent 
common for all measured beam energies.

\section{$\phi$ meson production}
In the context of strange particle production, the $\phi$ meson is of 
particular interest. Its overall strangeness neutrality makes it 
insensitive to the 
strange chemical potential in hadro-statistical models. On the other hand, 
consisting of a strange and an anti-strange quark it should be more 
sensitive than kaons to strangeness enhancement if the number of available 
strange quarks is determined in a prehadronic stage of the collision.
Previously, an enhancement in $\phi$ production per pion at top SPS 
energy of about a 
factor of 3, comparing central Pb+Pb to p+p collisions at the same beam 
energy, has been reported \cite{na49_phi}. We have now extended the 
analysis to central Pb+Pb at $80 \, A \rm GeV$ and $40 \, A \rm GeV$.

\begin{figure}[b]
\begin{center}
\begin{minipage} {4cm}
\begin{center}
\epsfxsize=4cm
\epsfbox{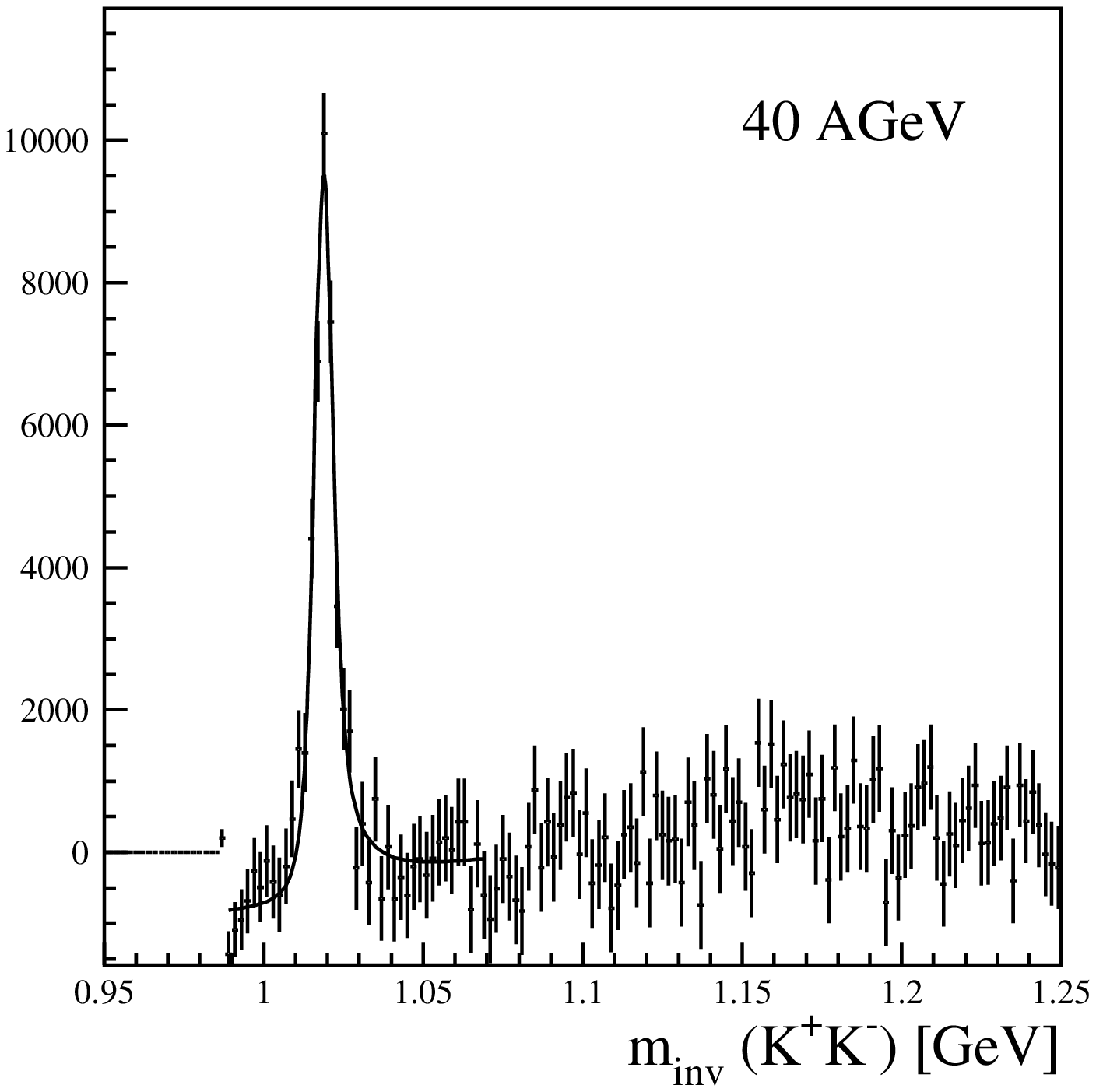}
\end{center}
\end{minipage}
\begin{minipage} {4cm}
\begin{center}
\epsfxsize=4cm
\epsfbox{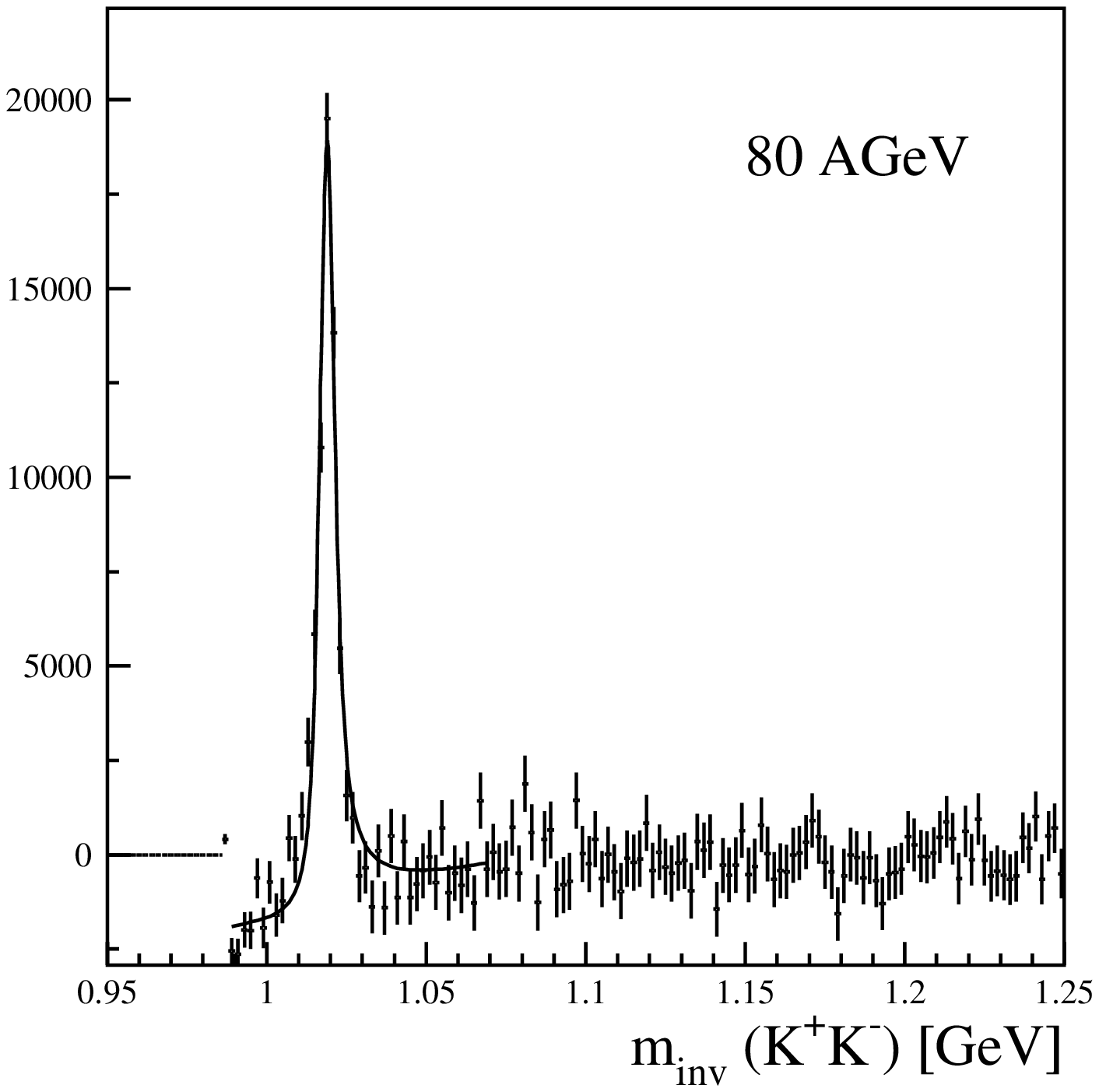}
\end{center}
\end{minipage}
\begin{minipage}{4cm}
\begin{center}
\epsfxsize=4cm
\epsfbox{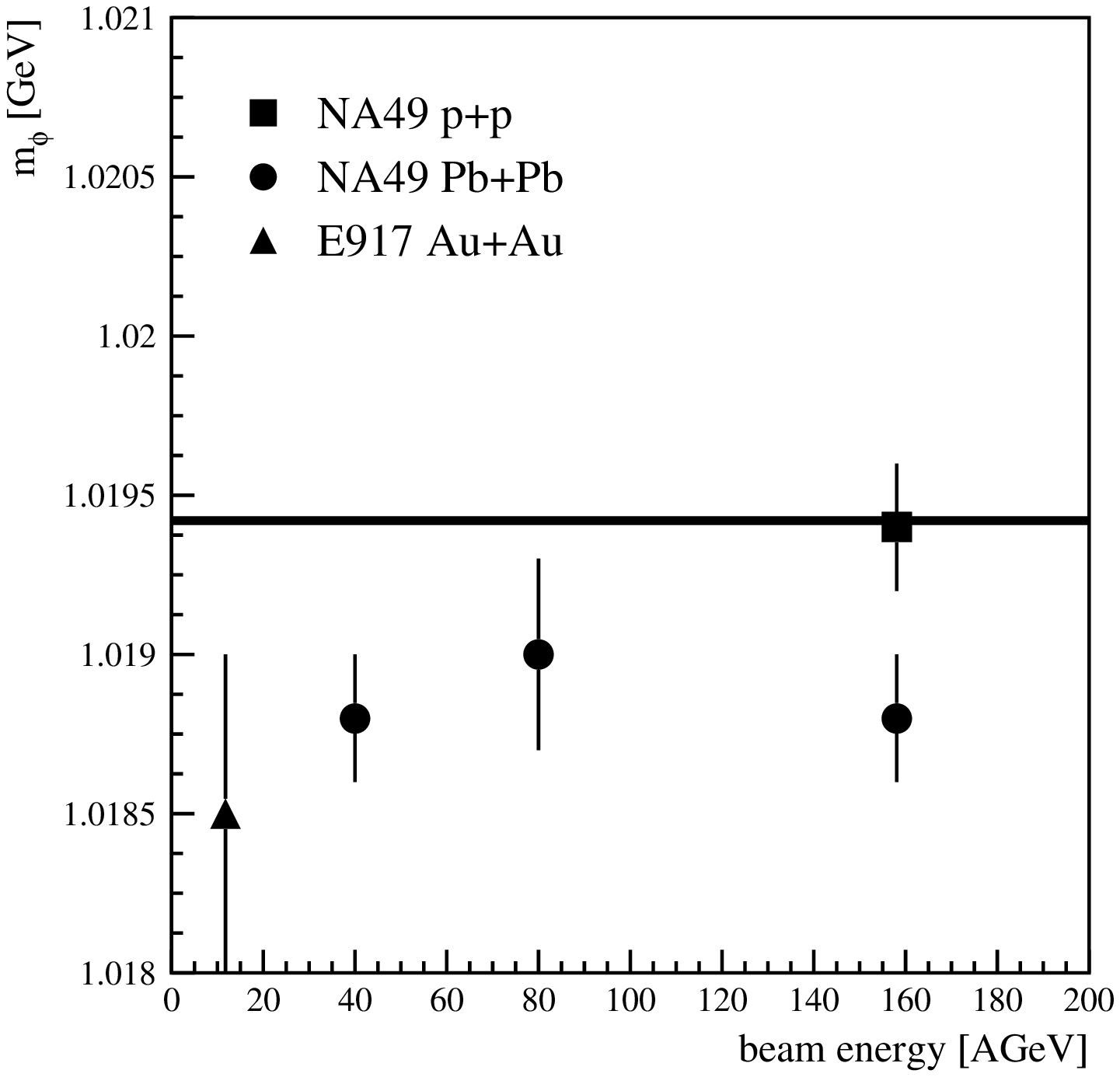}
\end{center}
\end{minipage}
\end{center}
\vspace{-0.5cm}
\caption{\label{fig:phi_minv} Invariant $K^+K^-$ mass in the forward 
hemisphere for $40 \, A \rm GeV$ (left) and $80 \, A \rm GeV$ (centre). 
The full lines represent fits of a relativistic Breit-Wigner distribution 
on top of a linear background in the vicinity of the signal. Right: $\phi$ 
mass as a function of beam energy. The full line represents the literature
value \cite{pdg}.}
\end{figure}

\begin{figure}[b]
\begin{center}
\begin{minipage} {5cm}
\begin{center}
\epsfxsize=5cm
\epsfbox{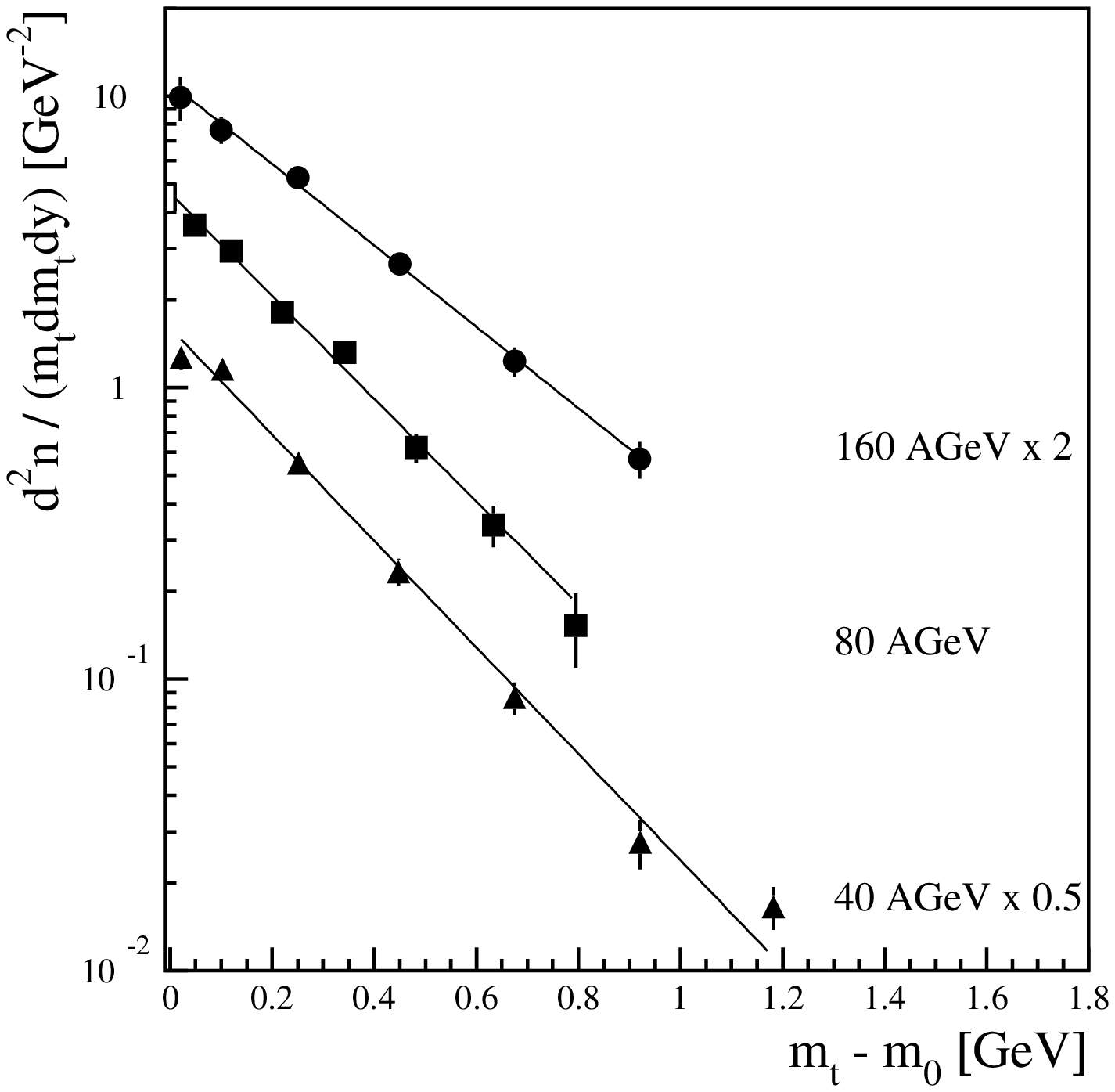}
\end{center}
\end{minipage}
\begin{minipage} {5cm}
\begin{center}
\epsfxsize=5cm
\epsfbox{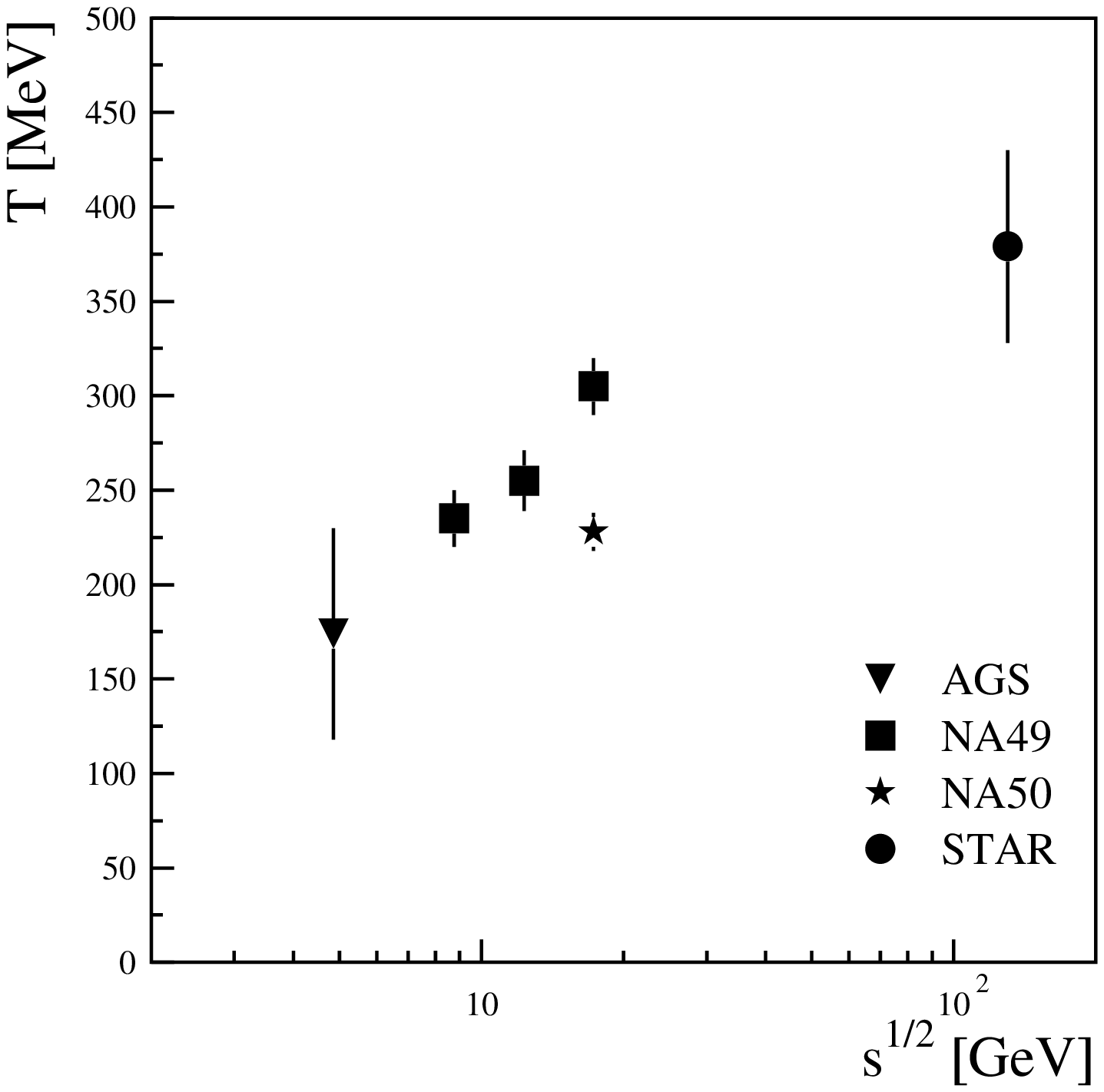}
\end{center}
\end{minipage}
\end{center}
\vspace{-0.5cm}
\caption{\label{fig:phi_transverse} Left: $m_t$ spectra of the $\phi$ meson. 
The spectra at different beam energies are scaled for better visibility. 
Right: $\phi$ slope parameter as a function of collision energy. Data not
measured by NA49 are taken from 
\cite{ags_phi}\cite{na50_phi}\cite{star_phi}.}
\end{figure}

In NA49, the $\phi$ meson is measured by its decay into charged kaons. 
Details of the analysis method can be found in \cite{na49_phi}. Figure 
\ref{fig:phi_minv} shows the invariant-mass signals at the two energies in 
the forward rapidity hemisphere. The depletion on the left side of the 
signal can be shown to stem from the final state strong interaction of $K^+$ 
and $K^-$. The widths of the $\phi$ peaks are consistent with the
free-particle width \cite{pdg} folded with the experimental resolution of 
about $1 \, \rm MeV$. For the peak positions, we find at all three energies 
analysed so far a slight ($\approx 0.5 \, \rm MeV$) deviation from the 
literature value \cite{pdg}, as depicted in the right panel of figure 
\ref{fig:phi_minv}. A similar deviation has been observed at the AGS 
\cite{ags_phi}, albeit with a larger error. In contrast, our analysis of 
p+p collisions at $158 \, A \rm GeV$ \cite{na49_phi} gives exactly the 
book value. It is still under investigation whether this shift is an 
experimental effect, possibly linked to the distortion in the 
background-subtracted invariant-mass spectra.

The transverse spectra of the $\phi$ can be seen in figure 
\ref{fig:phi_transverse}. Again, they are reasonably well described by 
exponential distributions, but in contrast to kaons, the slope parameter 
increases with beam energy (right panel of figure \ref{fig:phi_transverse}). 
It should be noted, however, that NA50 reports 
a different $T$ at $158 \, A \rm GeV$ \cite{na50_phi}.

\begin{figure}
\begin{center}
\epsfxsize=10cm
\epsfbox{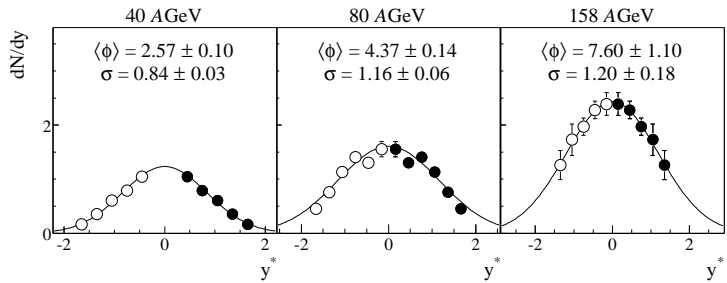}
\vspace{-0.5cm}
\caption{\label{fig:phi_rapidity} $\phi$ rapidity distributions. Solid 
symbols are measured data, open symbols are reflected at midrapidity. 
The full lines represent Gaussian fits. Left: $40 \, A \rm GeV$; centre: 
$80 \, A \rm GeV$; right: $158 \, A \rm GeV$.}
\end{center}
\end{figure}

\begin{figure}
\begin{center}
\begin{minipage} {5cm}
\begin{center}
\epsfxsize=5cm
\epsfbox{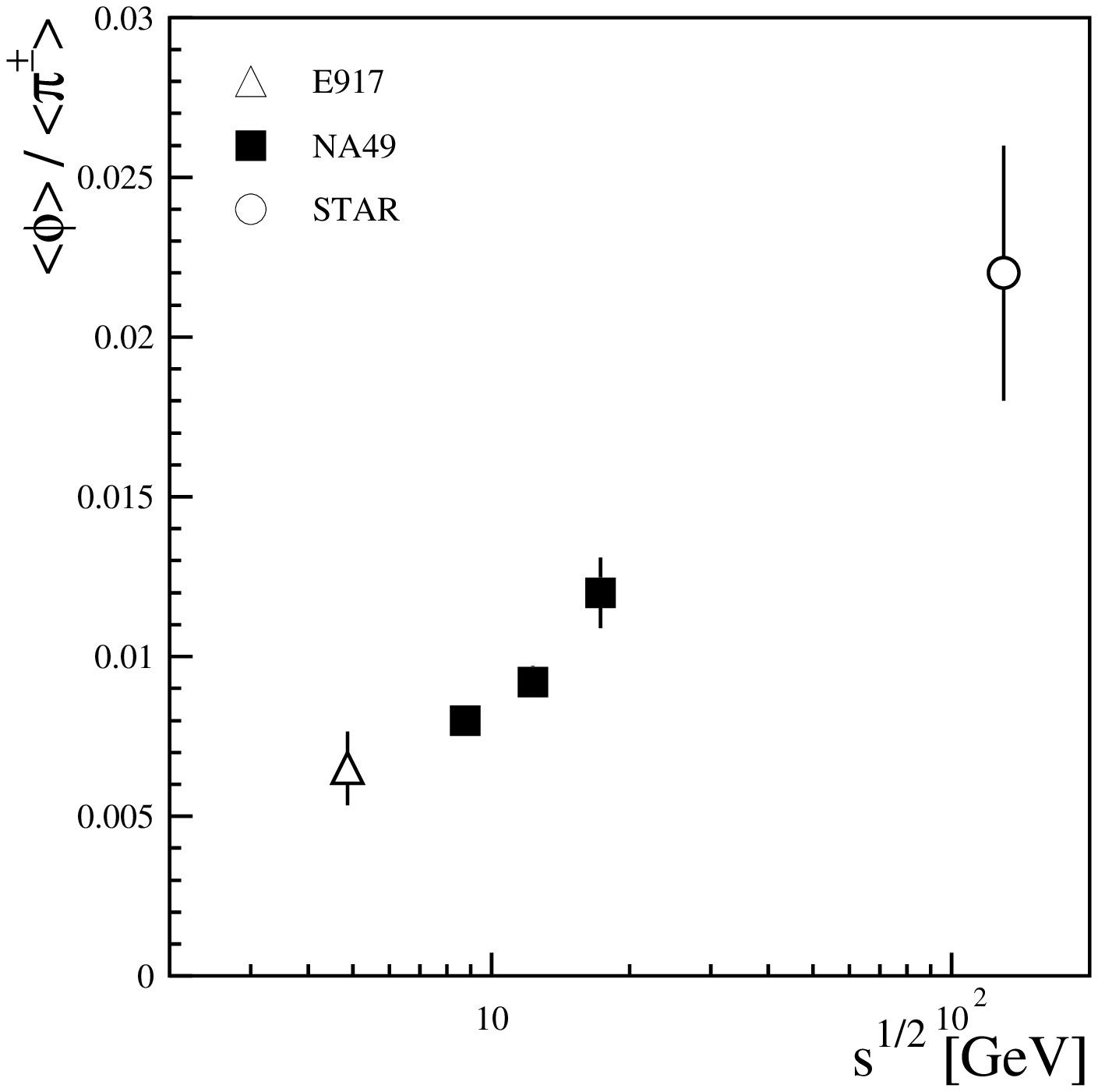}
\end{center}
\end{minipage}
\begin{minipage} {5cm}
\begin{center}
\epsfxsize=5cm
\epsfbox{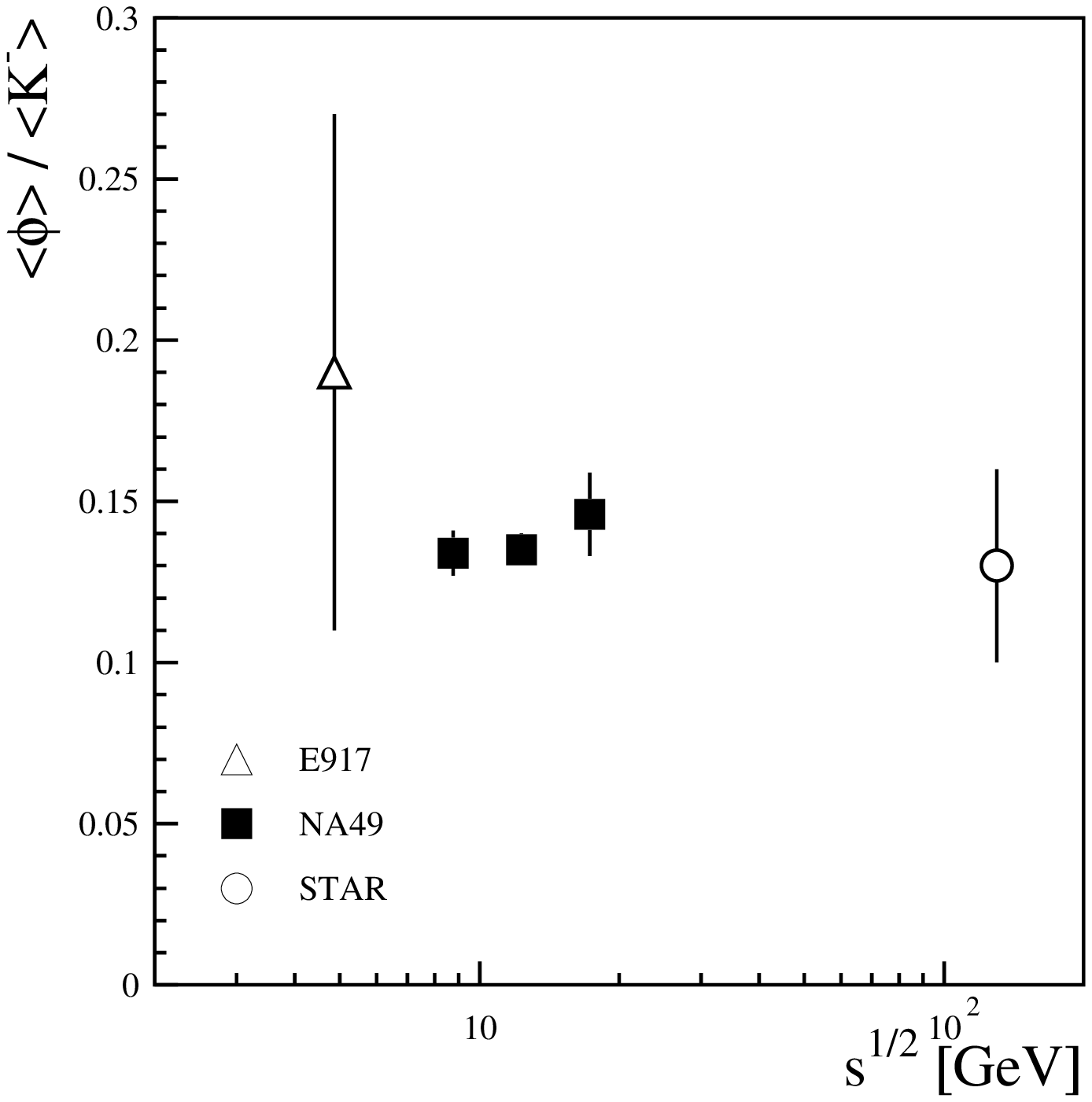}
\end{center}
\end{minipage}
\end{center}
\vspace{-0.5cm}
\caption{\label{fig:phi_yield} The ratios of $\phi/\pi^{\pm}$ (left) and $\phi/K^-$ 
(right) as functions of collision energy. For the data not measured by
NA49, see \cite{ags_phi}\cite{star_phi}\cite{energy_paper} and 
references therein.}
\end{figure}

Figure \ref{fig:phi_rapidity} shows the measured rapidity distributions of 
the $\phi$ meson at $40$ and $80 \, A \rm GeV$ in comparison to 
the published spectrum at top SPS energy. In all cases, a Gaussian 
distribution gives a good description of the data, with a width increasing 
with beam energy. This trend is in line with observations for pion and 
kaon production. Summing the measured bins and extrapolating to full phase 
space using the Gaussian fit, we obtained the total $\phi$ yield, 
which is shown
in figure \ref{fig:phi_yield} normalised to the average of the $\pi^+$ and 
$\pi^-$ yields and to the $K^-$ yield. We observe a monotonic increase 
in $\phi/\pi$ 
from AGS via SPS to RHIC energies, quite similar to the excitation function 
of $K^- / \pi^-$. This similarity can be straightforwardly seen in 
the $\phi/K^-$ ratio, which is, within errors, independent of $\sqrt{s}$.

\begin{figure}[t]
\begin{center}
\begin{minipage} {6cm}
\begin{center}
\epsfxsize=6cm
\epsfbox{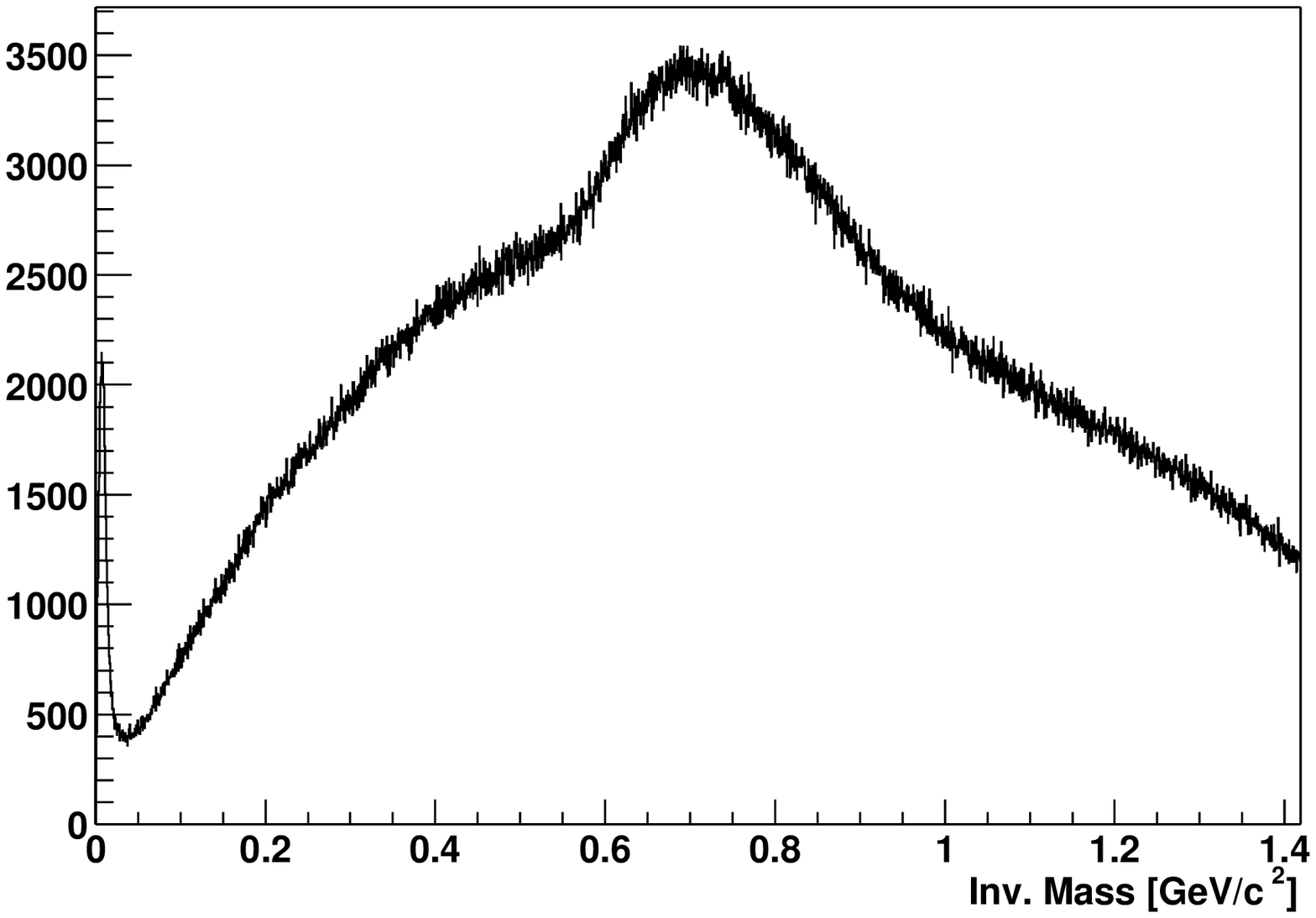}
\end{center}
\end{minipage}
\begin{minipage} {6cm}
\begin{center}
\epsfxsize=6cm
\epsfbox{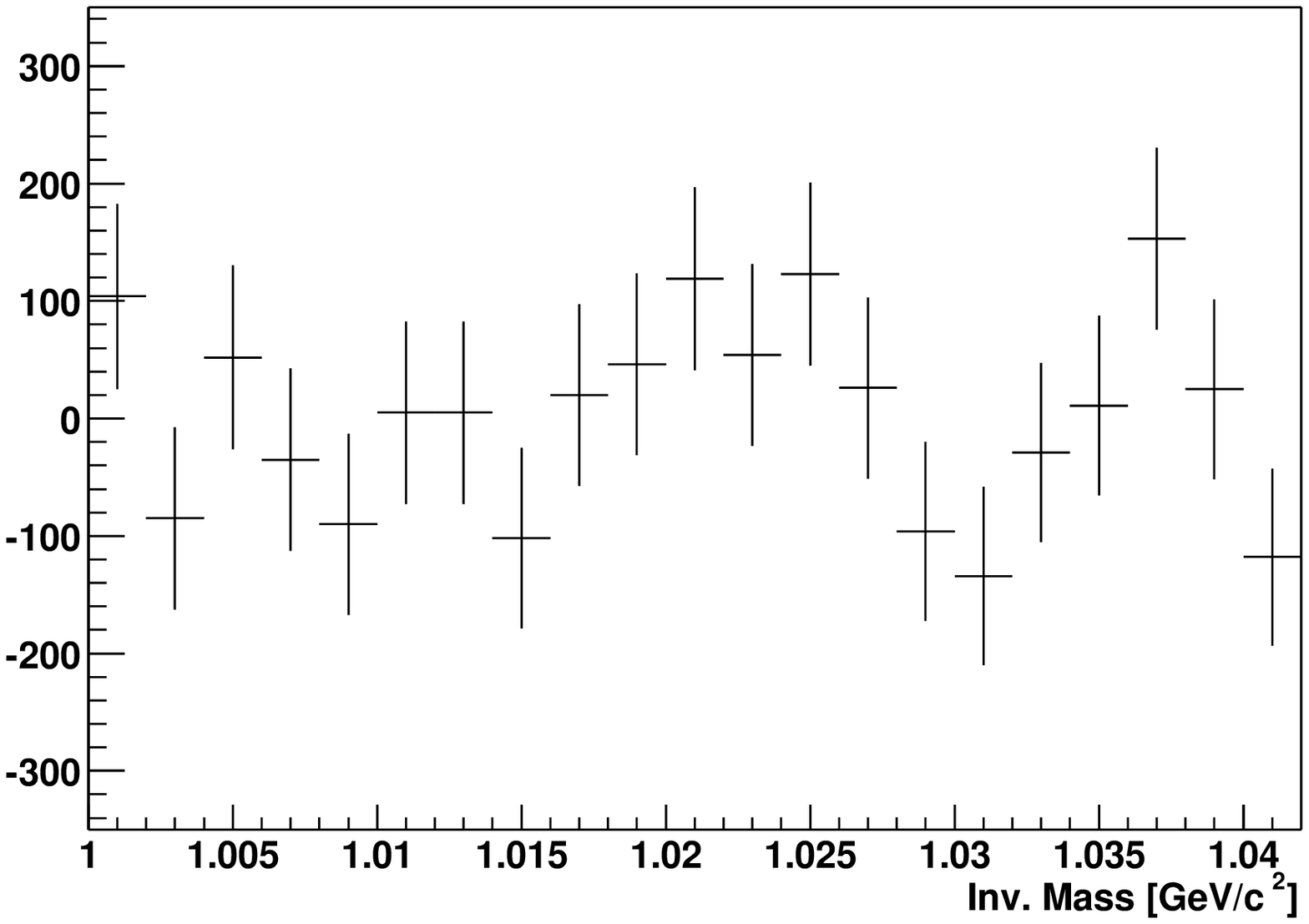}
\end{center}
\end{minipage}
\end{center}
\vspace{-0.5cm}
\caption{\label{fig:phi_ee} Left: $e^+e^-$ invariant-mass spectrum; right: 
invariant-mass spectrum after background subtraction in the $\phi$ mass 
region.}
\end{figure}

In the light of the discrepancies between the $\phi$ measurements in the 
$K^+K^-$ and $\mu^+\mu^-$ channels \cite{na49_phi,na50_phi}, both in slope 
parameter and total yield, NA49 has undertaken an effort to reconstruct 
the $\phi$ signal in the $e^+e^-$ decay channel, which should be similar to 
the decay into $\mu^+\mu^-$. For this analysis, the high-statistics data set 
at $158 \, A \rm GeV$ has been used. As figure \ref{fig:phi_ee} 
demonstrates, we do not observe a signal in this channel. Using the 
expected invariant-mass resolution obtained from a detailed detector
simulation,
we estimate the upper limit resulting from this analysis to be 
$\langle \phi \rangle <\approx 40$, with currently large systematic errors. Most probably, this 
measurement will unfortunately have no discriminating power to clear up 
the reported experimental differences.

\section{Chemical and kinetic freeze-out}
With a still growing wealth of data on strange particle production, the 
question arises whether these data give rise to a consistent picture of the 
reaction dynamics in heavy-ion collisions. Particle yields provide 
information about the chemical freeze-out stage of the collision, while 
the transverse spectra give insight into the kinetic freeze-out, i.~e.~the 
stage when elastic interactions cease.

\begin{figure}[t]
\begin{center}
\begin{minipage} {6.5cm}
\begin{center}
\epsfxsize=6.5cm
\epsfbox{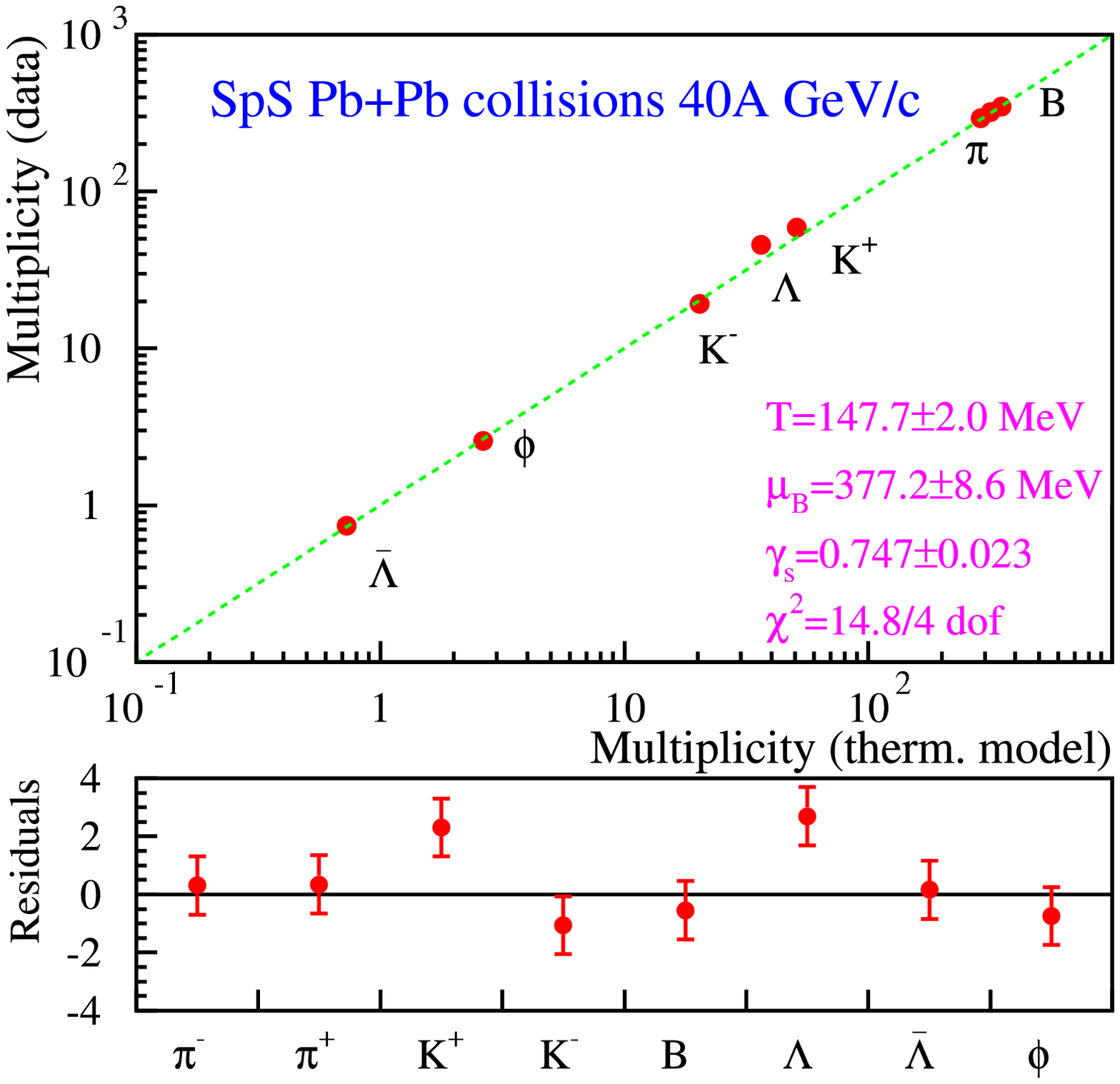}
\end{center}
\end{minipage}
\begin{minipage} {6.5cm}
\begin{center}
\epsfxsize=6.5cm
\epsfbox{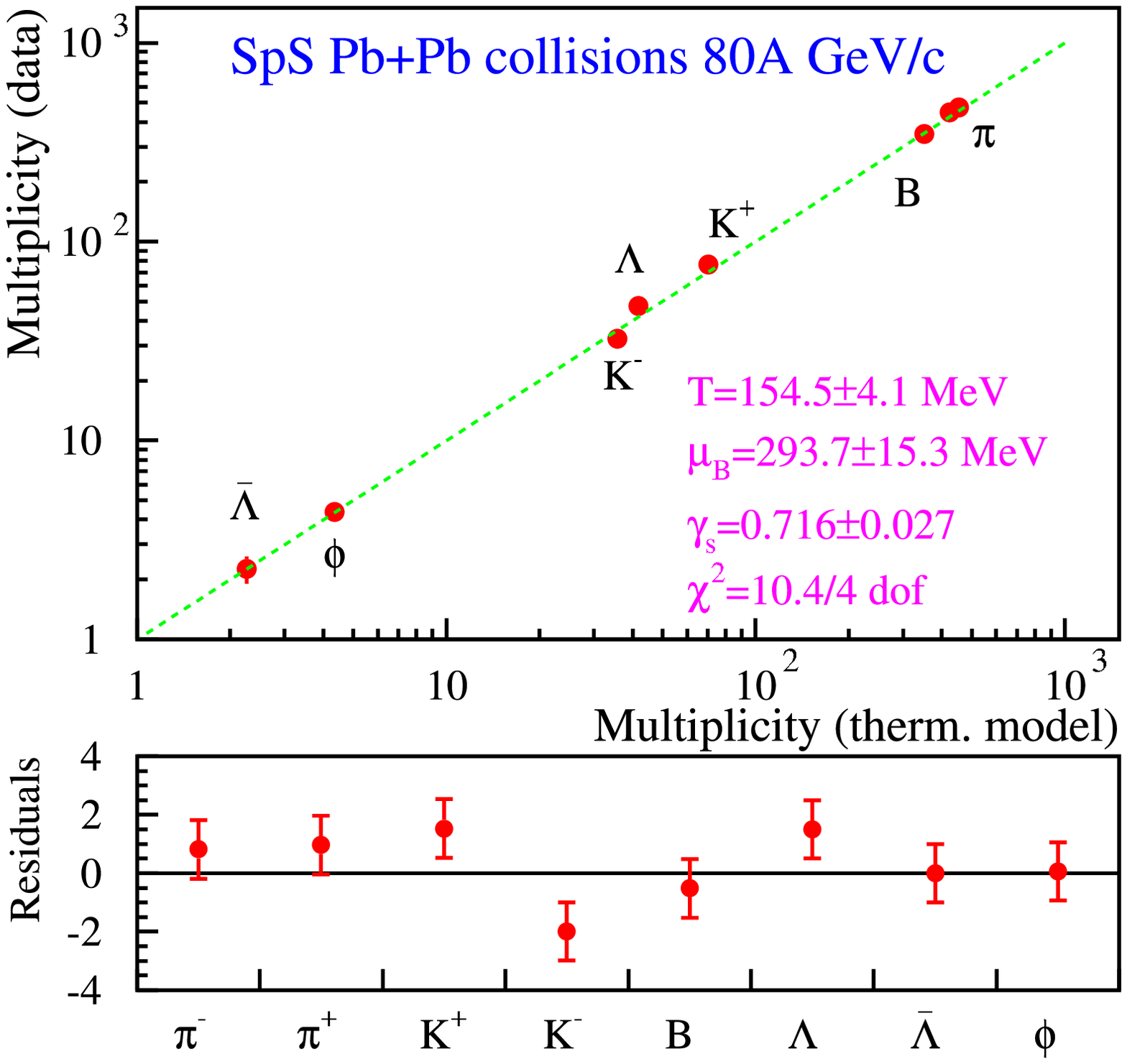}
\end{center}
\end{minipage}
\end{center}
\vspace{-0.5cm}
\caption{\label{fig:becattini} Comparison of the thermal model fit 
\cite{becattini,becat_private} with measured particle yields 
at $40 \, A \rm GeV$ (left) and $80 \, A \rm GeV$ (right).}
\end{figure}

\begin{figure}[b]
\begin{center}
\epsfxsize=6cm
\epsfbox{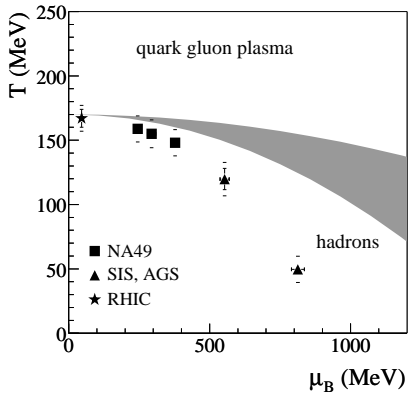}
\caption{\label{fig:freezeout} Chemical freeze-out points in the $T-\mu_B$ 
plane from the hadron gas model fits \cite{becat_private,cleymans}. 
The shaded region 
represents the phase boundary given in \cite{allton}.}
\end{center}
\vspace{-0.5cm}
\end{figure}

It has been shown before that the measured particle yields or yield ratios 
resulting from Pb+Pb collisions at top SPS energies can be reproduced by 
statistical models based on
a grand canonical ensemble with hadronic degrees of freedom. We thus employ 
a representative of this kind of models \cite{becattini} to fit the NA49 
data measured so far at $40$, $80$ and $158 \, A \rm GeV$. Note that the 
model used employs a strangeness suppression factor $\gamma_s$, which accounts
for an incomplete saturation of strangeness with respect to a fully 
equilibrated hadron gas. 
This additional parameter enables the model to reproduce the yields
also at the lower beam energies with 
reasonable accuracy as figure \ref{fig:becattini} demonstrates.
The  fit parameters, namely the temperature $T$, the baryochemical 
potential $\mu_B$ and $\gamma_s$ reflect the
conditions at chemical freeze-out. Figure \ref{fig:freezeout} shows the 
results in the $T$-$\mu_B$ plane together with results obtained for SIS, 
AGS and RHIC energies. It appears that chemical freeze-out occurs on a 
smooth curve in this plane which has been interpreted as being determined 
by a constant energy per particle \cite{cleymans}. Moreover, at SPS the 
freeze-out curve approaches the expected phase boundary between 
hadronic matter and the Quark-Gluon-Plasma as calculated with lattice QCD 
\cite{allton}.

\begin{figure}[b]
\begin{center}
\begin{minipage} {7cm}
\begin{center}
\epsfxsize=7cm
\epsfbox{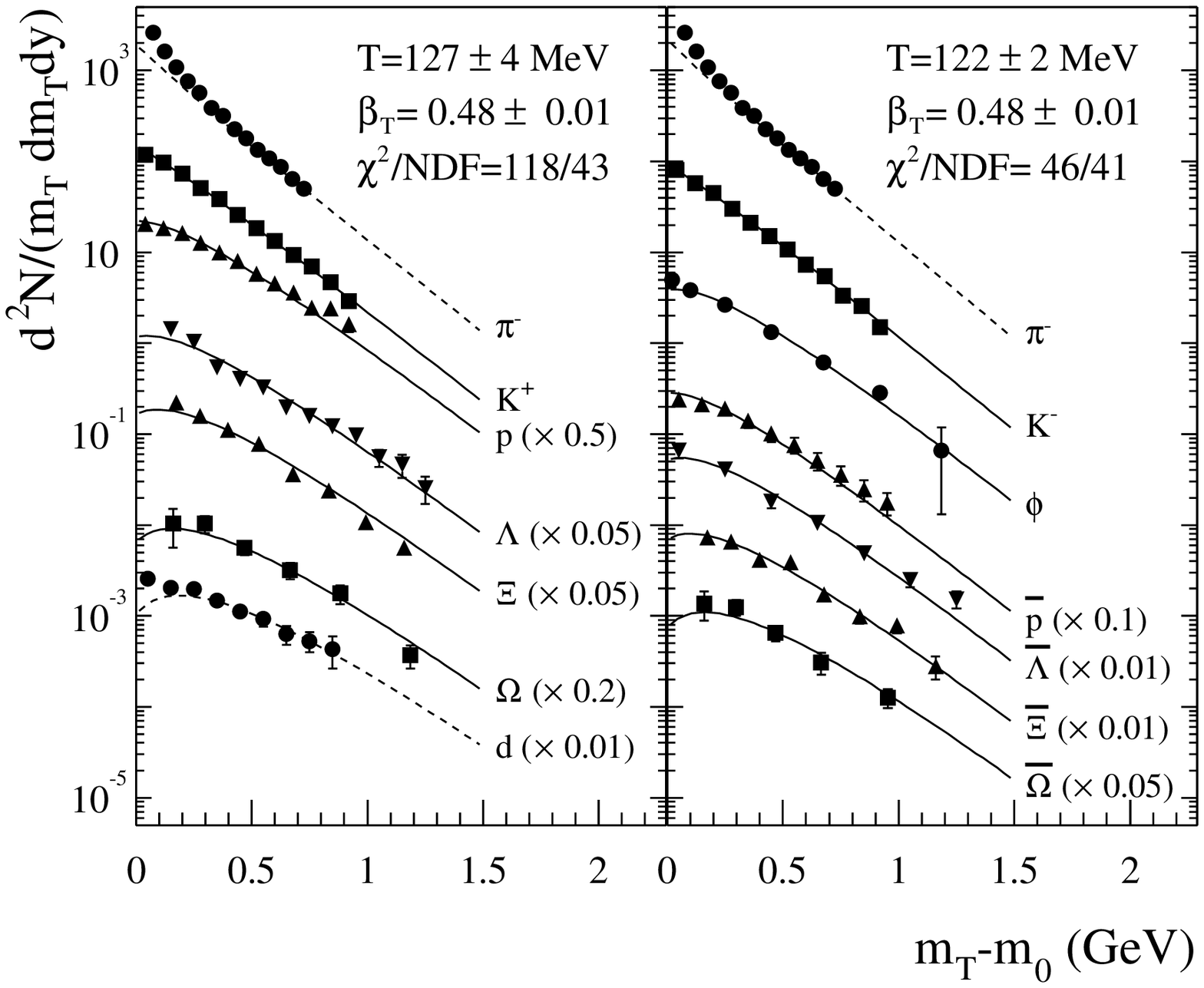}
\end{center}
\end{minipage}
\begin{minipage} {7cm}
\begin{center}
\epsfxsize=7cm
\epsfbox{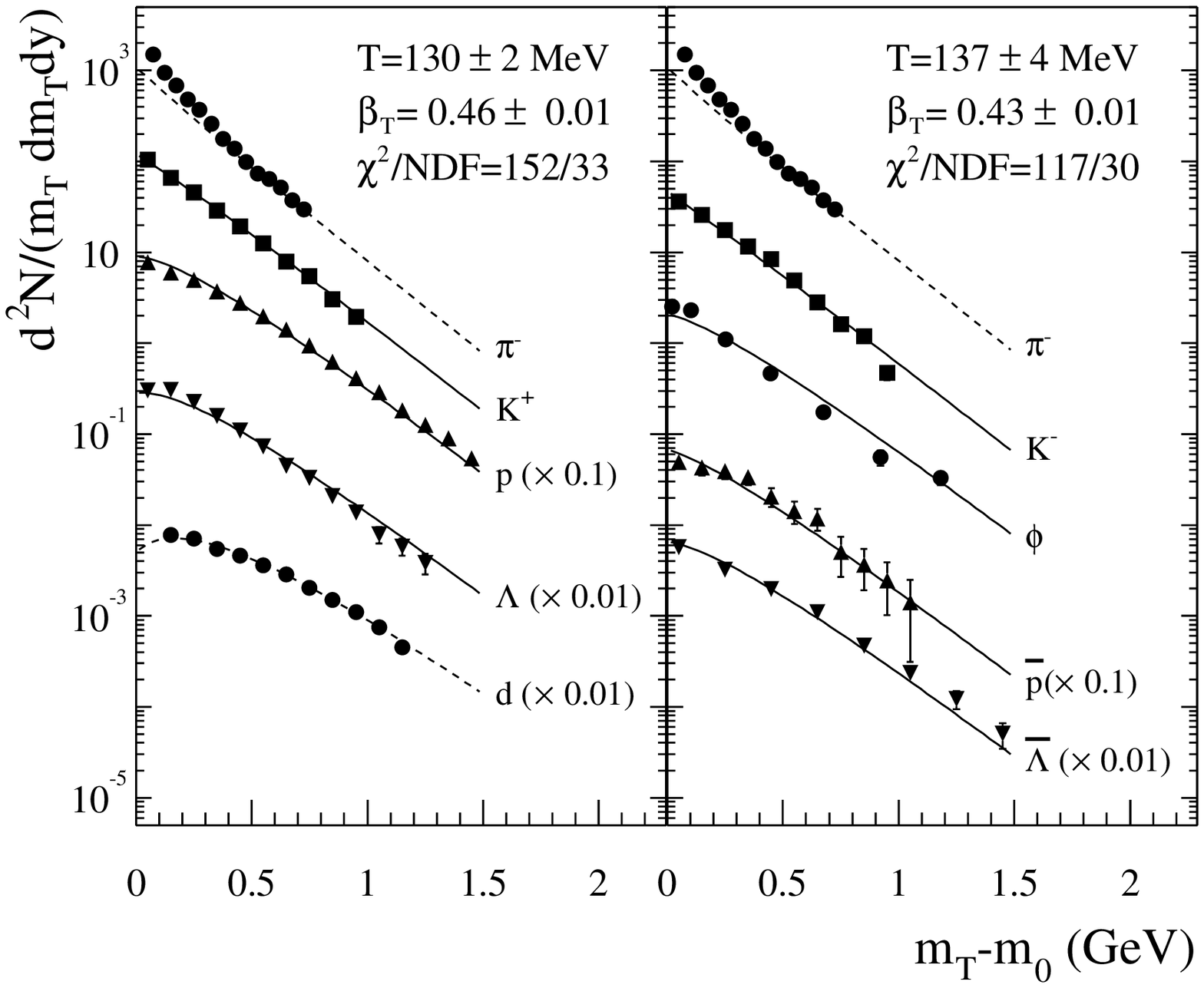}
\end{center}
\end{minipage}
\end{center}
\vspace{-0.5cm}
\caption{\label{fig:blastwave} Blast wave fits \cite{schneder} to the 
transverse spectra measured by NA49. Pions and deuterons were excluded 
from the fits. Left: $158 \, A \rm GeV$; right: $40 \, A \rm GeV$.}
\end{figure}

Figure \ref{fig:blastwave} depicts a compilation of hadron transverse mass
spectra, measured by NA49 at $40 \, A \rm GeV$ and $158 \, A \rm GeV$. 
The lines represent a radial-flow fit \cite{schneder}, assuming a constant 
transverse expansion velocity and common kinetic freeze-out for all 
particle types. Pions have been excluded from the fit because of the
significant contribution of resonance decays to their low-$p_t$ yields. 
Obviously, the model can well describe all data at both energies, which 
is also true for $80 \, A \rm GeV$. The values of the fit parameters, 
temperature $T$ and expansion velocity $\beta_T$, are similar for all three 
energies. Although the model used may employ oversimplified assumptions, 
it shows that the available data can be understood in the framework
of a locally equilibrated, rapidly expanding fireball. We observe no 
deviations of the heavy hyperons ($\Xi,\Omega$) from this picture, which 
would signal an earlier freeze-out for these particles.

\section{Summary}
We have presented new experimental results obtained within the 
energy scan programme of NA49, in particular kaon production at 
$30 \, A \rm GeV$ and $\phi$ production 
at $40 \, A \rm GeV$ and $80 \, A \rm GeV$. The excitation function of the 
$K^+/\pi^+$ ratio shows a sharp maximum around $30 \, A \rm GeV$, which has 
been predicted by the Statistical Model of the Early Stage, thus supporting 
the scenario of a phase transition at low SPS energies. 
The observation that the kaon slope parameters do not depend on beam 
energy in the SPS regime supplies further evidence for 
this picture. The slope of the $\phi$ meson, in contrast, rises with beam 
energy. The $\phi/K^-$ ratio is found to be independent of collision energy; 
the $\phi/\pi$ ratio thus increases smoothly from AGS via SPS to RHIC. The 
measured particle yields can be well reproduced by a hadron gas model 
employing a strangeness undersaturation factor $\gamma_S$. Chemical freeze-out 
seems to occur on a smooth curve in the $T-\mu_B$ plane, close to the 
expected phase boundary to the QGP. The transverse spectra can be 
interpreted at the three energies analysed so far by a rapidly expanding
thermal source, showing no evidence of an earlier decoupling of the heavy 
hyperons.

\ack{ \small
This work was supported by the Director, Office of Energy Research, 
Division of Nuclear Physics of the Office of High Energy and Nuclear 
Physics of the US Department of Energy (DE-ACO3-76SFOOO98 and 
DE-FG02-91ER40609), the US National Science Foundation, 
the Bundesministerium fur Bildung und Forschung, Germany, 
the Alexander von Humboldt Foundation, 
the UK Engineering and Physical Sciences Research Council, 
the Polish State Committee for Scientific Research (2 P03B 130 23, 
SPB/CERN/P-03/Dz 446/2002-2004, 2 P03B 02418, 2 P03B 04123), 
the Hungarian Scientific Research Foundation (T032648, T14920 and T32293),
Hungarian National Science Foundation, OTKA, (F034707),
the EC Marie Curie Foundation,
and the Polish-German Foundation.
}



\section*{References}
\begin{thebibliography}{20}
\bibitem{stock} Stock R 1999 \PL B {\bf 456} 277--282
\bibitem{na49_lambda} Mischke A for the NA49 collaboration 2002 
\jpg {\bf 28} 1761--1768
\bibitem{sqm03_meurer} Meurer C for the NA49 collaboration 
{\it These proceedings}
\bibitem{sqm03_mitrov} Mitrovski M for the NA49 collaboration 
{\it These proceedings}
\bibitem{nim} Afanasiev S V \etal [NA49 collaboration] 1999 \NIM A 
{\bf 430} 210-244
\bibitem{energy_paper} Afanasiev S V \etal [NA49 collaboration] 2002 
\PR C {\bf 66} 054902
\bibitem{gorgad03} Gorenstein M I, Ga{\'z}dzicki  M and Bugaev K 2003 
{\it Preprint} hep-ph/0303041
\bibitem{pbm} Braun-Munzinger P \etal 2002 \NP A {\bf 697} 902--912
\bibitem{rqmd} Sorge H, St\"ocker H and Greiner W 1989 \NP A
{\bf 498} 567
\bibitem{urqmd} Bass S A \etal [UrQMD collaboration] 1998
{\it Prog.~Part.~Nucl.~Phys.} {\bf 41} 255-369
\bibitem{SMES} Ga{\'z}dzicki M and Gorenstein M I 1999 
{\it Acta Phys.~Polon.} B {\bf 30} 2705--2735
\bibitem{na49_phi} Afanasiev S V \etal [NA49 collaboration] 2000 \PL B 
{\bf 491} 59--66
\bibitem{pdg} Hagiwara K \etal [Particle Data Group] 2002 \PR D 
{\bf 66} 010001
\bibitem{ags_phi} Seto R K and Xiang H for the E917 collaboration 1999
\NP A {\bf 661} 506--509
\bibitem{na50_phi} Alessandro B \etal [NA50 collaboration] 2003 \PL B 
{\bf 555} 147--155
\bibitem{star_phi} Laue F for the STAR collaboration 2002 \jpg 
{\bf 28} 2051--2057
\bibitem{becattini} Becattini F, Ga{\'z}dzicki M and Sollfrank J 1998 
{\it Eur.~Phys.~J.} C {\bf 5} 143--153
\bibitem{becat_private} Becattini F 2003 {\it Private communication}
\bibitem{cleymans} Becattini F \etal 2001 \PR C {\bf 64} 024901
\bibitem{allton} Allton C R \etal 2002 {\it Preprint} hep-lat/0204010
\bibitem{schneder} Schnedermann E and Heinz U 1994 \PR C {\bf 50} 1675--1683





\end{thebibliography}
\end{document}